\documentclass[usenatbib,letterpaper,hyperref]{mn2e}
\usepackage[totalwidth=480pt, totalheight=680pt]{geometry}
\usepackage{times}
\usepackage{epsfig}
\usepackage{amsmath}
\usepackage{amssymb}
\usepackage{url}
\usepackage{aas_macros}
\usepackage[labelfont=bf,labelsep=period]{caption}
\usepackage{longtable}

\title[Diversity of dwarf galaxy rotation curves]{The unexpected diversity of dwarf galaxy rotation curves}
\author[K. A. Oman et al.]{\newauthor Kyle
  A. Oman$^{1,}$\thanks{koman@uvic.ca}, Julio F. Navarro$^{1,2}$,
  Azadeh Fattahi$^{1}$, Carlos S. Frenk$^{3}$, \newauthor Till Sawala$^{3}$, Simon D. M. White$^{4}$,
  Richard Bower$^{3}$, Robert A. Crain$^{5}$, \newauthor Michelle Furlong$^{3}$,
  Matthieu Schaller$^{3}$, Joop Schaye$^{6}$, Tom Theuns$^{3}$
\\
$^{1}$ Department of Physics \& Astronomy, University of Victoria, Victoria, BC, V8P 5C2, Canada\\
$^{2}$ Senior CIfAR Fellow\\
$^{3}$ Institute for Computational Cosmology, Department of Physics, University of Durham, South Road, Durham DH1 3LE, United Kingdom\\
$^{4}$ Max-Planck Institute for Astrophysics, Garching, Germany\\
$^{5}$ Astrophysics Research Institute, Liverpool John Moores University, IC2, Liverpool Science Park, 146 Brownlow Hill, Liverpool, L3 5RF, United Kingdom\\
$^{6}$ Leiden Observatory, Leiden University, PO Box 9513, NL-2300 RA Leiden, the Netherlands
}
\date{\today}

\def\gtrsim{\mathrel{\raise0.35ex\hbox{$\scriptstyle >$}\kern-0.6em
\lower0.40ex\hbox{{$\scriptstyle \sim$}}}}
\def\lesssim{\mathrel{\raise0.35ex\hbox{$\scriptstyle <$}\kern-0.6em
\lower0.40ex\hbox{{$\scriptstyle \sim$}}}}

\begin{document}
\label{firstpage}
\maketitle

\begin{abstract} 
  We examine the circular velocity profiles of galaxies in
  $\Lambda$CDM cosmological hydrodynamical simulations from the
  {\small EAGLE} and {\small LOCAL GROUPS} projects and compare them
  with a compilation of observed rotation curves of galaxies spanning
  a wide range in mass. The shape of the circular velocity profiles of
  simulated galaxies varies systematically as a function of galaxy
  mass, but shows remarkably little variation at fixed maximum
  circular velocity. This is especially true for low-mass dark
  matter-dominated systems, reflecting the expected similarity of the
  underlying cold dark matter haloes. This is at odds with observed
  dwarf galaxies, which show a large diversity of rotation curve
  shapes, even at fixed maximum rotation speed. Some dwarfs have
  rotation curves that agree well with simulations, others do not. The
  latter are systems where the inferred mass enclosed in the inner
  regions is much lower than expected for cold dark matter haloes and
  include many galaxies where previous work claims the presence of a
  constant density ``core''.  The ``cusp vs core'' issue is thus better
  characterized as an ``inner mass deficit'' problem than as a density
  slope mismatch. For several galaxies the magnitude of this inner
  mass deficit is well in excess of that reported in recent
  simulations where cores result from baryon-induced fluctuations in
  the gravitational potential. We conclude that one or more of the
  following statements must be true: (i) the dark matter is more
  complex than envisaged by any current model; (ii) current
  simulations fail to reproduce the diversity in the effects of baryons on the inner
  regions of dwarf galaxies; and/or (iii) the mass profiles of ``inner
  mass deficit'' galaxies inferred from kinematic data are incorrect.
\end{abstract}


\section{Introduction}
\label{SecIntro}

Cosmological simulations have led to a detailed theoretical
characterization of the clustering of dark matter on galaxy scales. It
is now well established that, when baryons may be neglected, the
equilibrium mass profiles of cold dark matter (CDM) haloes are
approximately self-similar and can be adequately approximated by a
simple formula \citep{Navarro1996,Navarro1997}. The ``NFW profile'',
as this is commonly known, has a formally divergent density ``cusp''
near the centre, $\rho \propto r^{-\gamma}$, with $\gamma=1$, and
steepens gradually at larger radii. The corresponding circular
velocity profile, $V_{\rm circ}(r)$, is thus relatively steep near the centre,
$V_{\rm circ}\propto r^{1/2}$, in contrast with the rotation curves of some
dwarf galaxies, where the inner rotation speed rises linearly with
radius. The latter behaviour suggests that the dark matter density
profile has a shallower inner slope than predicted by simulations,
closer to a constant density ``core'' rather than a steeply divergent
``cusp''. This ``cusp vs core'' problem
\citep{Moore1994,Flores1994} has been known since the mid 1990s and
has elicited a number of proposed solutions.

One is that the dark matter is not ``cold''. Cores can be produced in
dark matter haloes by particle physics effects if the dark matter
particles have specific properties that differ from those of
WIMPs or axions, the standard CDM candidates. For example, phase space
constraints give rise to cores in “warm” dark matter (WDM) haloes
\citep[e.g.][]{BodeOstrikerTurok2001,Lovell2012}, although current
lower limits on WDM particle masses imply cores that are much smaller
than those inferred for many dwarfs
\citep{Maccio2012,Shao2013,Viel2013}. 

Alternatively, elastic collisions between particles of
“self-interacting” dark matter \citep[SIDM; see,
e.g.,][]{SpergelSteinhardt2000} may create cores, provided the
cross-sections are the right size. SIDM has fallen somewhat out of
favour because of concerns that it may fail to account for the central
dark matter density profiles of galaxy clusters
\citep{Miralda-Escude2002} or that it would lead to the dissolution of
individual galaxies in clusters \citep{GnedinOstriker2001}. However,
recent work has concluded that those arguments may be countered by
appealing to velocity-dependent cross sections
\citep{Vogelsberger2012} or by re-evaluating carefully the
observational constraints. Indeed, \citet{Rocha2013} argue that a
velocity-independent specific cross-section of order
$\sigma/m \sim 0.1$~cm$^2$~g$^{-1}$ can reproduce simultaneously dwarf
and cluster observations, although this conclusion relies on a
relatively uncertain extrapolation of their results to the regime of
dwarfs. As a result, the situation remains
unsettled. \citet{Zavala2013}, for example, have recently argued
that only a finely-tuned SIDM model can be reconciled with
observation while \citet{Elbert2014} have concluded the opposite,
although we note that the latter work is based on resimulations of
only two halos of similar ($\sim 40$ km/s) circular velocity.

An alternative is that rotation curves may be reconciled with $\Lambda$CDM
haloes by ``baryon effects'' operating during the formation of the
galaxy \citep{NavarroEkeFrenk1996,GnedinZhao2002,Read2005}. In
particular, recent simulations in which star formation occurs in a
series of short bursts where dense clouds of gas are continually
assembled and violently dispersed have been shown to drive potential
fluctuations that can induce constant density cores at the centre of
$\Lambda$CDM haloes
\citep{MashchenkoCouchmanWadsley2006,Governato2010,Governato2012,Brook2012,Teyssier2013,Madau2014}. Although
there is no consensus that galaxy formation necessarily has this
effect \citep[other simulations make realistic galaxies without
producing cores; see, e.g.,][and our discussion below]{Schaller2014},
there is growing consensus that the inner dark matter profiles can, in
principle, be reshaped during the formation of a galaxy, even a dark
matter-dominated one, in a manner that may depend on its merger
history \citep[see, e.g.,][]{diCintio2014,Onorbe2015}. ``Baryon-induced'' cores in
CDM halos would be difficult to distinguish from those
produced by other effects, such as collisional effects in the case of
SIDM \citep[see, e.g.,][]{Vogelsberger2014a,BastidasFry2015},
complicating matters further.

\begin{figure}
{\leavevmode \includegraphics[width=\columnwidth]{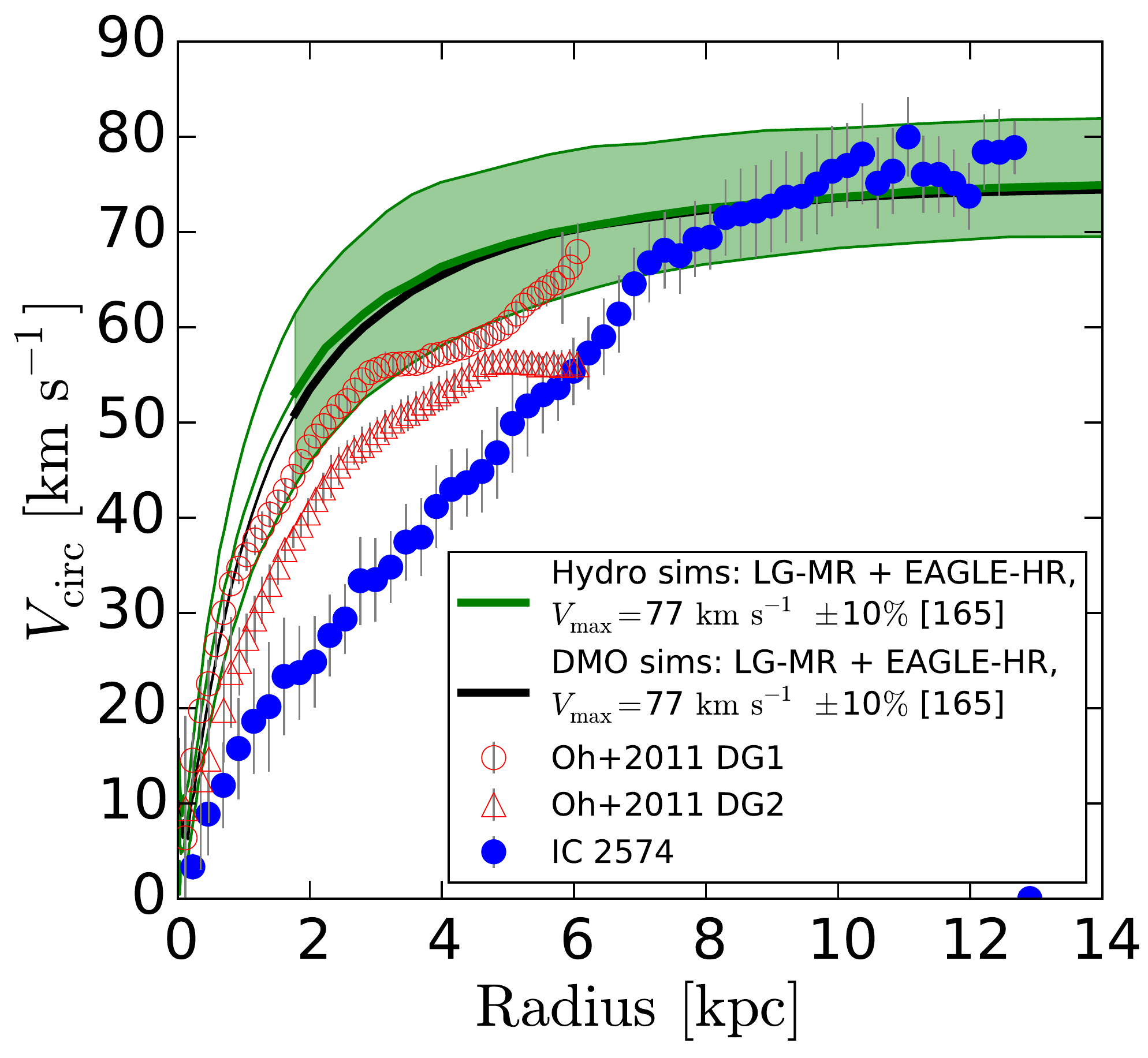}}
\caption{Rotation curves of IC~2574 (filled circles) and of the
  simulated galaxies DG1 (open circles) and DG2 (open triangles), taken from \citet{Oh2011}.
  The green line shows the median circular
  velocity curve of all galaxies
  from our LG-MR and EAGLE-HR simulations (see \S\ref{SecNumSims}) with $V_{\rm max}=77$~km~s$^{-1}\pm10$ per cent, matching the value of
  $V_{\rm max}$ of IC~2574. The shaded area indicates the $10^{\rm th}$--$90^{\rm th}$ percentile range at
  each radius. The lines become thinner and the shading stops inside
  the average convergence radius, computed following the prescription
  of \citet{Power2003}. The numbers in square brackets in the legend are the numbers of
  galaxies/haloes that contribute to that velocity bin.
  The solid black line is the median circular velocity profile of
  haloes of the same $V_{\rm max}$, identified in our dark
  matter-only simulations.
  \label{FigDG1}}
\end{figure}

Finally, it has been argued that a critical reappraisal of the actual
constraints placed on the inner dark matter density slope by
observations might be needed, citing concerns about complexities such as
non-circular motions \citep{Swaters2003}, instrumental smearing
effects \citep{Swaters2003,vandenBosch2000}, and/or departures from
axisymmetry \citep{Hayashi2004}, all of which may complicate the
relation between the observational data and the underlying
gravitational potential. The advent of two-dimensional observational
surveys
with better sensitivity and angular resolution have allayed some of
these earlier concerns \citep[e.g.,][]{Simon2003,deBlok2008,KuziodeNaray2008,Oh2011,Adams2014,Oh2015}, and have led to the view that reliable
determination of the inner slope of the dark matter density profile is
indeed possible.

Some questions, however, remain. \citet{Adams2014}, for example,
report inner slopes as steep as $\gamma=0.67\pm 0.10$
for a sample of seven nearby dwarfs, whereas \citet{Oh2011} report much shallower
slopes ($\gamma=0.29\pm 0.07$) for seven dwarfs selected from the
THINGS survey \citep{Walter2008}. Whether these discrepancies reflect
a genuine physical difference between the galaxies in each of those
samples, or a systematic difference in the modeling of the
observational data, is still unclear. What is clear is that the
inferred slopes are highly sensitive to how the mass of the baryonic
population is modeled as well as to how the inevitable presence of
noncircular motions near the centre is accounted for.

The case of NGC~2976 offers a sobering example: when inner kinematic
peculiarities in the gas are ignored a nearly constant density core is
inferred \citep{Simon2003}, while a much steeper slope is inferred
from Jeans modeling of stellar tracers \citep{Adams2012}. Although the
disagreement can be resolved once the non-circular motions are
accounted for and the total mass of the stellar component is better
constrained \citep{Adams2014}, this example illustrates the difficulty
of inferring $\gamma$, even when quality multi-tracer data at high resolution
are available.

\begin{center}
\begin{figure*}
{\leavevmode \includegraphics[width=2.\columnwidth]{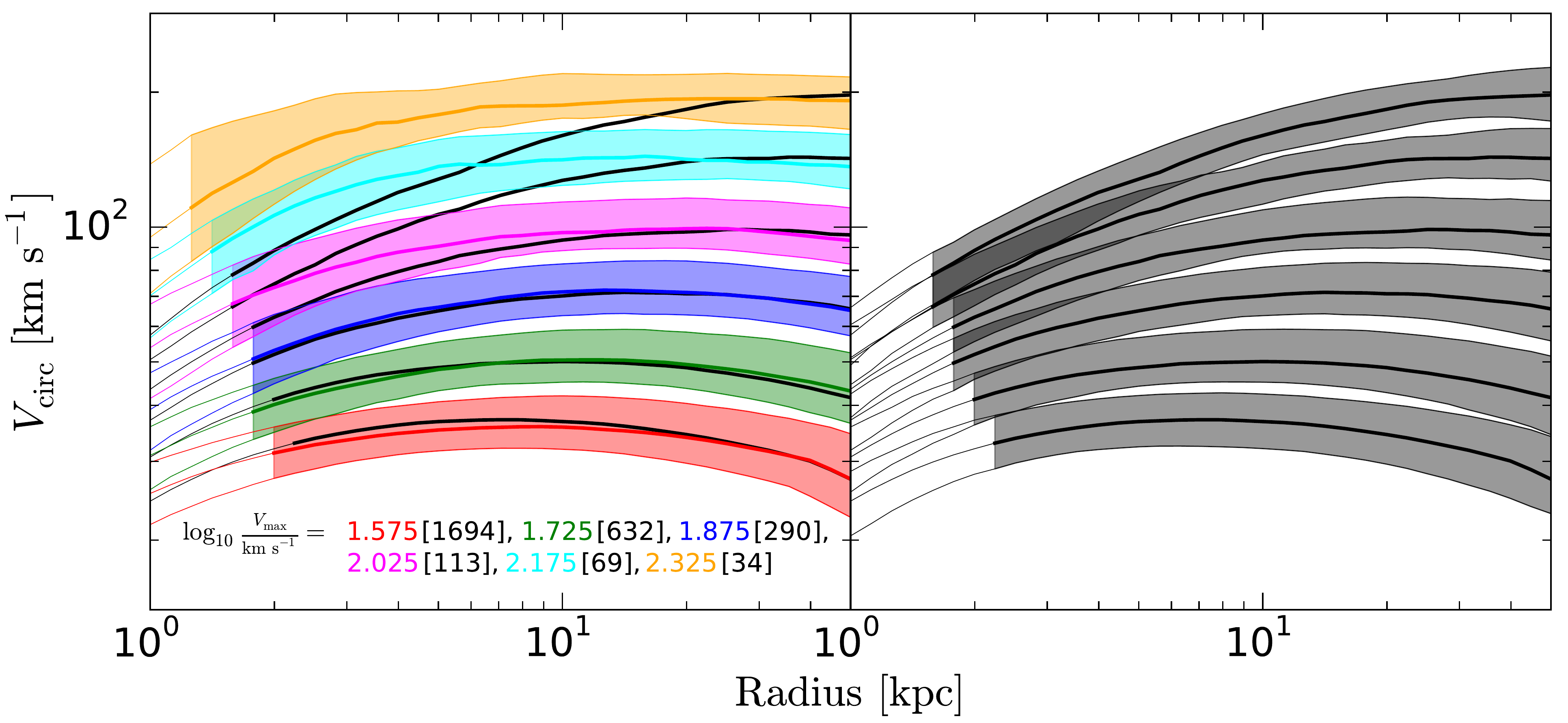}}
\caption{Circular velocity curves of simulated galaxies in the {\small
    EAGLE-HR} and {\small LG-MR} simulations, which have similar mass
  and force resolution. The left panel shows the results for
  hydrodynamical simulations; the right panel shows results for the
  corresponding dark matter-only (DMO) runs. Systems are grouped
  according to their maximum circular velocity in bins of $0.15$-dex
  width as listed in the legend. The number of systems in each bin is
  also listed in the legend, in square brackets. Solid curves indicate
  the median circular velocity curve for galaxies in each bin; the
  shaded areas show the $10^{\rm th}$--$90^{\rm th}$ percentile
  range. The curves become thinner line and shading stops inside the
  average convergence radius, computed for each bin using the
  prescription of \citet{Power2003}. The solid black curves are the
  same in both panels, and indicate the median circular velocity
  curves in the DMO simulations.  \label{FigSimRotCurLog}}
\end{figure*}
\end{center}

A further drawback of focusing the discussion on the central value of
$\gamma$ is that it risks missing an important dimension of the
problem, which concerns the total mass enclosed within the inner
regions of a galaxy. This may be illustrated by the case of IC~2574,
one of seven systems whose rotation curves were compared by
\citet{Oh2011} to simulated galaxies where baryon-induced fluctuations
had flattened the dark matter cusp \citep{Governato2010}. Oh et
al. conclude that the simulated galaxies have ``haloes with a central
mass distribution similar to that observed in nearby dwarf galaxies'',
a conclusion based on the similarity of the innermost values of
$\gamma$. 

Although the value of $\gamma$ in the inner $\sim$1~kpc of these
systems might be similar, the full circular velocity profiles of
IC~2574 and simulated galaxies are actually very different. We show
this in Fig.~\ref{FigDG1}, where we plot the circular velocity profile
of IC~2574 (filled circles) with that of DG1 (open circles) and DG2
(open triangles), the two simulated galaxies from the Oh et al. study.
The simulated galaxies show a clear excess of mass in the inner
regions compared to IC~2574, despite the ``cores'' in the dark matter
carved out by baryons. The reason for the discrepancy is that these
cores are small, and only affect the inner kpc, whereas IC~2574 shows
a linearly-rising rotation curve out to $\sim 8$ kpc. The
``baryon-induced'' cores in these simulations are clearly too small to
reconcile CDM with rotation curves of galaxies like IC~2574, so the
reported agreement between observation and simulation is, in this
case, illusory.

The above discussion demonstrates that resolving the ``cusp vs core''
problem requires more than just matching the innermost values of
$\gamma$. Even if baryon effects are able to flatten the innermost
value of $\gamma$ to values consistent with observed estimates
\citep[see, e.g., fig.~2 in][for a comparison at $500$ pc from the
centre]{Pontzen2014}, this is not enough to ensure that simulated
rotation curves agree with observation. We argue therefore for a
reassessment of the ``cusp vs core'' controversy where full circular
velocity profiles of observed galaxies are directly compared with the
results of cosmological hydrodynamical simulations. 
This has only
become possible very recently, given the advent of cosmological
hydrodynamical simulations able to produce a realistic galaxy
population and, presumably, also realistic rotation curves \citep[see,
e.g.,][]{Vogelsberger2014b,Schaye2015}. 

The analysis we advocate here, which extends to lower masses that of
\citet{Schaller2014}, has a number of advantages. One is that the
inner regions, which are difficult to observe {\it and} to simulate,
are less emphasized in the comparison. The second is that it makes
full use of the predictive power of the $\Lambda$CDM paradigm. Earlier
work had left considerable room for discussion because of
uncertainties, for example, in the normalization of the NFW
mass-concentration relation, which determines the actual density
profile of a halo of given mass. That debate has now been settled: the
cosmological parameters are known to exquisite accuracy \citep[see,
e.g.,][]{Planck2015}, and large cosmological simulations with
excellent resolution have characterized conclusively the
mass-concentration relation, its normalization, and scatter
\citep[see,
e.g.,][]{Neto2007,Duffy2008,Maccio2008,Zhao2009,Prada2012,Ludlow2014}. As
a result, simulated galaxies can now be compared directly with
observations without need for rescalings or other adjustments.

We adopt this view here by considering the circular velocity profiles
of galaxies selected from the {\small EAGLE} and {\small LOCAL GROUPS}
simulation projects. The simulated galaxies cover a wide range of
maximum circular velocity, from $25$ to $250$~km~s$^{-1}$, and are compared with data
compiled from the literature for galaxies that span a similar range in
maximum rotation velocity. We begin by presenting the simulated curves in
Sec.~\ref{SecSimRotCur} and the observed compilation in
Sec.~\ref{SecObsRotCur}. We then compare them and discuss our
results in Sec.~\ref{SecCompRotCur}, before summarizing our main
conclusions in Sec.~\ref{SecConc}.

\begin{center}
\begin{figure}
\leavevmode \includegraphics[width=\columnwidth]{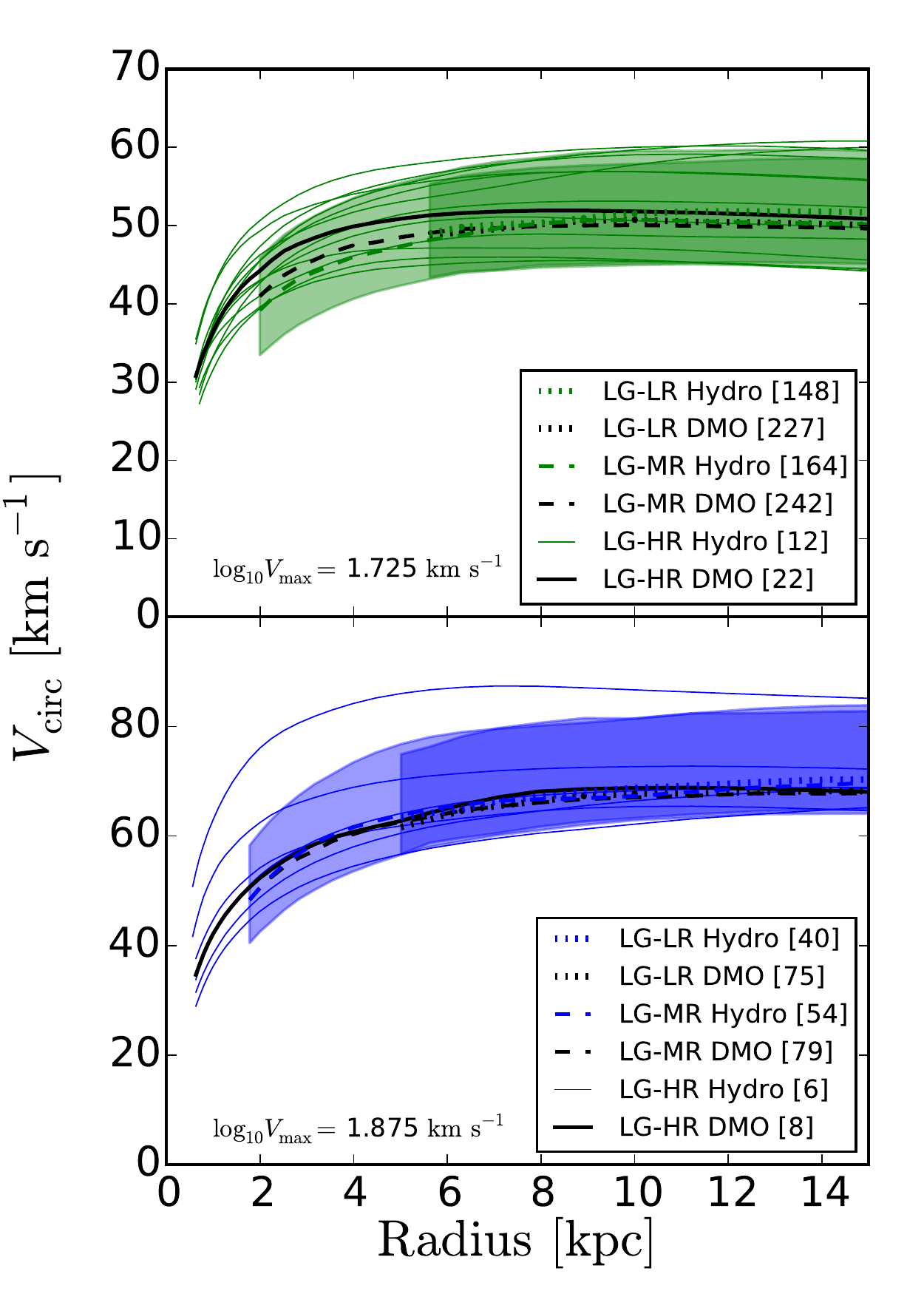}
\caption{As Fig.~\ref{FigSimRotCurLog}, but for only two velocity
  bins, and shown in linear units. The total number of galaxies in
  each bin are listed in the legend. The dotted lines correspond to
  galaxies in the low-resolution LG-LR simulations, shown only down to their
  convergence radii. The dashed line shows the same, but for
  galaxies in the medium resolution LG-MR simulations. Thin solid curves correspond
  to individual systems in the high-resolution LG-HR simulations. Only two of
  the twelve LG volumes have been simulated at high resolution, hence the
  relatively small number of individual curves in each panel. The
  black lines show the median circular velocity profile for dark matter-only simulations. \label{FigSimRotCurLin}}
\end{figure}
\end{center}

\section{Circular Velocity Profiles of Simulated Galaxies}
\label{SecSimRotCur}

We discuss here the circular velocity curves of galaxies selected from
the {\small EAGLE} \citep{Schaye2015,Crain2015} and {\small LOCAL
  GROUPS} \citep{Sawala2015} simulation projects. {\small EAGLE} is
calibrated to reproduce, in a cosmological volume, the observed population of
galaxies, including their abundance as a function of galaxy mass and
their typical size. {\small LOCAL GROUPS} simulates volumes tailored
to match the Local Group environment with the same physics as {\small
  EAGLE} but at higher numerical resolution. We refer the interested
reader to the above references for details, and provide here only a
brief summary of the parameters most relevant to our analysis.

\subsection{The numerical  simulations}
\label{SecNumSims}

\subsubsection{The {\small EAGLE} simulations}
\label{SecEAGLE}

The {\small EAGLE} project is a suite of cosmological hydrodynamical
simulations run with a substantially modified version of the {\small
  P-GADGET3} code, itself a descendent of the publicly available
{\small GADGET2} \citep{Springel2005}. In this work we use only the highest-resolution
realization in the suite, which we denote EAGLE-HR
\citep[Ref-L025N0752\footnote{Though the fiducial {\small EAGLE} model for this box size and resolution is Recal-L025N0752, we use the Ref-L025N0752 model because it more closely matches the subgrid physics used in the {\small LOCAL GROUPS} simulations.} in the nomenclature of][]{Schaye2015}. This simulation
has a cube side length of $25$ comoving Mpc; $752^3$ dark
matter particles each of mass $1.21\times10^{6}\,{\rm M}_\odot$; the
same number of gas particles each of mass $2.26\times10^{5}\,{\rm
  M}_\odot$; and a Plummer-equivalent gravitational softening length of $350$ proper pc
(switching to comoving above redshift 2.8). The cosmology adopted is that of the
\citet{Planck2014}, with $\Omega_m = 0.307$, $\Omega_\Lambda = 0.693$,
$\Omega_b = 0.04825$, $h = 0.6777$ and $\sigma_8 = 0.8288$. {\small
  EAGLE} uses the pressure-entropy formulation of smoothed particle
hydrodynamics \citep{Hopkins2013}, and includes subgrid models for
radiative cooling, star formation, stellar and chemical enrichment,
energetic stellar feedback, black hole accretion and mergers, and AGN
feedback. 

\subsubsection{The {\small LOCAL GROUPS} simulations}
\label{SecLG}

The {\small LOCAL GROUPS} project is a suite of zoom-in hydrodynamical simulations of regions selected from cosmological dark matter-only simulations to contain two haloes with approximately the masses and dynamics of the Milky Way and M31, and no other nearby large structures (Fattahi et al 2015, in preparation). {\small LOCAL GROUPS} uses the same code and physics as the ``Ref'' {\small EAGLE} simulations. The same volumes are simulated at three different resolution levels, which we denote LG-LR, LG-MR and LG-HR for low, medium and high resolution, respectively. Each resolution level is separated by a factor of $\sim 10$ in particle mass and a factor of $\sim 2$ in force resolution.

Table~\ref{LGparams} summarizes the particle masses and softening lengths of each resolution level. We note that the LG-MR resolution level corresponds closely to EAGLE-HR. There are $12$ volumes, all of which have been simulated at LG-LR and LG-MR resolution levels. Only volumes $1$ and $4$ have been simulated at high resolution. For each hydrodynamical simulation there is also a corresponding dark matter-only (DMO) simulation of the same region at the same resolution. The WMAP7 cosmological parameters \citep{WMAP7} are used in {\small LOCAL GROUPS}\footnote{The differences between the cosmological parameters used in the {\small EAGLE} and {\small LOCAL GROUPS} projects are very small and of little consequence to this study.}, with $\Omega_m = 0.2727$, $\Omega_\Lambda = 0.728$, $\Omega_b = 0.04557$, $h = 0.702$ and $\sigma_8 = 0.807$.

The {\small LOCAL GROUPS} simulation suite produces realistic Local Group-like environments, reproducing the stellar mass function of Milky Way and M31 satellites, and that of Local Group dwarf galaxies \citep{Sawala2014} using the same calibration parameter choices as the {\small EAGLE}-Ref simulations.

\begin{table}
  \caption{Summary of the key parameters of the {\small EAGLE} and {\small LOCAL GROUPS} simulations used in this work. Details of the cosmological parameters are available in \citet[][WMAP7]{WMAP7} and \citet[][Planck13]{Planck2014}.\label{LGparams}}
  \begin{tabular}{lllll}
    \hline
    \hline
    & \multicolumn{2}{l}{Particle Masses (${\rm M}_\odot$)} & Max Softening & \\
    Simulation & DM & Gas & Length (pc)& Cosmology\\
    \hline
    EAGLE-HR & $1.2\times10^6$ & $2.3\times10^5$ & $350$ & Planck13\\
    LG-LR & $7.3\times10^6$ & $1.5\times10^6$ & $712$ & WMAP7 \\
    LG-MR & $5.9\times10^5$ & $1.3\times10^5$ & $308$ & WMAP7 \\
    LG-HR & $5.0\times10^4$ & $1.0\times10^4$ & $134$ & WMAP7\\
    \hline
  \end{tabular}
\end{table}

\subsubsection{Halo finding}

Structures are identified in our simulations using the {\small SUBFIND} algorithm \citep{Springel2001,Dolag2009}. Particles are first grouped into friends-of-friends (FoF) haloes by linking together dark matter particles separated by less than 0.2 times the mean interparticle spacing; gas and star particles are assigned to the same FoF halo as their nearest dark matter particle. Substructures are then separated along saddle points in the density distribution; in this step dark matter, gas and star particles are treated as a single distribution of mass. Finally, particles that are not gravitationally bound to the substructures are removed. The result is a collection of substructures termed ``subhaloes'', each typically corresponding to a single galaxy.

\subsection{Circular velocity curves}

The circular velocity profiles of simulated galaxies, $V_{\rm circ}(r) =(GM(<r)/r)^{1/2}$,
where $M(<r)$ is the mass enclosed within radius $r$, are shown in the
left panel of Fig.~\ref{FigSimRotCurLog}. Here we have gathered all
``central'' galaxies (i.e. excluding satellites) in the 12 LG-MR simulations,
together with all centrals in the EAGLE-HR simulation, and binned them
according to their maximum circular velocity, $V_{\rm max}$. We
adopted bins of $0.15$-dex width, centred at $\log_{10} (V_{\rm max}$/km~s$^{-1})=1.575, 1.725, 1.875, 2.025, 2.175, 2.325$, and show the median rotation
curve for galaxies in each bin with solid lines. The shaded areas indicate, at each radius,
the $10^{\rm th}$ and $90^{\rm th}$ percentile velocity for all galaxies in each
bin. The number of galaxies in each bin is listed in the legend.

This figure illustrates two important points. One is that the shapes of the circular
velocity curves of $\Lambda$CDM galaxies are a strong function of the maximum
circular velocity of the system. Indeed, once $V_{\rm max}$ is
specified, the circular velocity profile of a system is very well
constrained at all radii that are resolved\footnote{We adopt in all cases the
  ``convergence radius'' introduced by \citet{Power2003}, as computed
  from the DMO simulations. This radius marks the point where curves turn
  thinner and the shading stops in all figures.} by
the simulations. The second point is that, in general, circular
velocity curves of systems with substantially different $V_{\rm max}$ do not cross,
so that in principle a well measured circular velocity at
almost any radius may be translated into an excellent constraint on
$V_{\rm max}$.

These characteristics of the circular velocity curves are largely a
reflection of the self-similar nature of $\Lambda$CDM haloes, modified
by the baryonic component. This may be seen in the right panel of
Fig.~\ref{FigSimRotCurLog}, which is analogous to that in the left,
but for systems identified in dark matter-only simulations
of the same {\small EAGLE} and {\small LOCAL GROUPS} volumes. (The
solid black lines are the same in both panels and indicate the median
rotation curves in the DMO simulations.)  As discussed by
\citet{Schaller2014}, the effects of baryons are mainly discernible in
systems with $V_{\rm max} > 60$~km~s$^{-1}$. In those systems, the assembly
of the baryonic component of the galaxy leads to an increase in 
mass that tends to flatten the $V_{\rm circ}(r)$ profile in the
inner few kpc. 

In systems with $V_{\rm max}< 60$~km~s$^{-1}$, on the other hand, the
galaxy formation ``efficiency'' is very low, and the baryonic mass of
the central galaxy has, in general, a negligible effect on the
circular velocity curve, even in the inner regions. Our simulations
thus show little evidence for the formation of a constant-density
``core'' near the centre of dwarf galaxies, suggesting that the dark
matter ``core creation'' mechanism discussed by \citet{Pontzen2014} is
not a general result of $\Lambda$CDM simulations that produce
realistic galaxy populations, but rather a consequence of particular
algorithmic choices adopted to simulate star formation and feedback.

Indeed, simulations that produce ``cores'' generally adopt a high
density threshold for star formation \citep[$n_{H}\gtrsim
100$~cm$^{-3}$, e.g.][]{Governato2010} that results in micro bursts of
star formation concentrated in highly compact gas clouds that can be
rapidly dispersed by feedback. This mode of star formation is not
present in our simulations, which adopt a lower effective star
formation threshold \citep[$n_{H}\gtrsim
0.1$~cm$^{-3}\left(Z/0.002\right)^{-0.64}$, depending on the
metallicity $Z$ and motivated by models of the ${\rm H\,I}$--H$_2$
transition;][]{Schaye2004} because we do not attempt to model a cold
($T << 10^{4}$~K) insterstellar gas phase. Our simulations thus allow
star formation to occur throughout the rotationally-supported gaseous
disk of a galaxy, limiting the sudden fluctuations in the
gravitational potential on small scales.

Although we do not see obvious evidence for constant density cores in
the circular velocity profiles, we do find $\sim 1$ kpc cores in the
dark matter density profile of some galaxies with $V_{\rm max}\gtrsim
100$~km~s$^{-1}$. These cores are {\it only} present in the dark
matter -- any dark mass displaced is actually replaced by baryons so
that the net result is typically an overall increase in the total mass
in the inner regions and a steepening of the potential. As a result, these
cores canonot explain the linearly-rising rotation curves of dwarf
galaxies, and are of little consequence to the rest of our analysis.

Fig.~\ref{FigSimRotCurLin} offers evidence that the lack of ``cores''
in the {\it total} mass profiles in our simulations is not a result of
insufficient numerical resolution. Here we show, in linear units, the
circular velocity profiles of {\small LOCAL GROUPS} galaxies in two
bins of $V_{\rm max}$, simulated at three different numerical
resolutions (LG-HR, LG-MR, and LG-LR; see Sec.~\ref{SecLG}). As in Fig.~\ref{FigSimRotCurLog},
the shaded regions in Fig.~\ref{FigSimRotCurLin} show the $10^{\rm
  th}$--$90^{\rm th}$ percentile range spanned by the $V_{\rm circ}(r)$
curves in each bin, for the medium-resolution (MR) and low-resolution
(LR) simulations. The thin lines in Fig.~\ref{FigSimRotCurLin}
correspond to individual systems identified in the high-resolution
(HR) version of the same simulations. Fig.~\ref{FigSimRotCurLin} shows that, at all
well-resolved radii, the circular velocity profiles are insensitive to
numerical resolution, despite the fact that the LG simulation series
span a factor of $12^2=144$ in particle mass and of more than $\sim 5$
in force resolution.

We emphasize that, although our simulations do not form ``cores'',
they do produce galaxies whose abundance, structural properties, and
evolution seem in good accord with observational constraints
\citep[see,
e.g.,][]{Schaye2015,Sawala2014,Furlong2015,Schaller2014}. The
formation of dark matter ``cores'' thus does not appear to be a
requisite ingredient of galaxy formation simulations that successfully
reproduce the structural properties of the observed galaxy population,
at least for galaxies with stellar masses $M_{*}\gtrsim
10^{9}$~M$_\odot$.

\begin{center}
\begin{figure*}
{\leavevmode \includegraphics[width=2.\columnwidth]{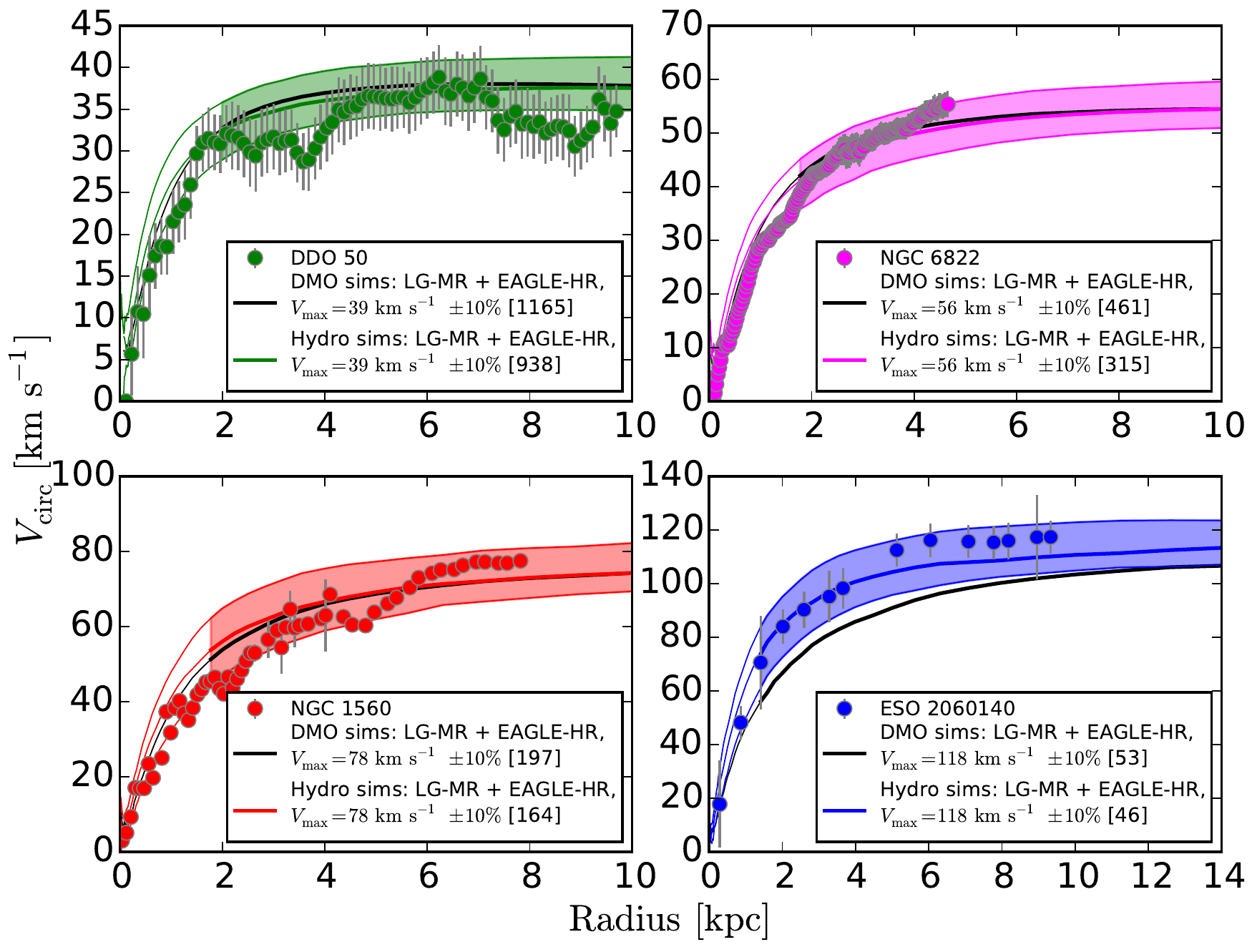}}
\caption{Four examples of galaxies in our sample with rotation curves that are in good agreement with the circular velocity curves of
  our $\Lambda$CDM hydrodynamical simulations. The four galaxies have been
  chosen to
  span a wide range in maximum circular velocity, from $\sim30$ (top left) to $\sim120$~km~s$^{-1}$ (bottom right). As in
  Fig.~\ref{FigSimRotCurLog}, the solid curves and shaded areas show
  the median (and $10^{\rm th}$--$90^{\rm th}$ percentile range) of all simulated
  galaxies in $20$ per cent-width bins centred at the maximum circular
  velocity of the galaxy in each panel (see legend). The solid black
  curve corresponds to the median circular velocity curve of our DMO simulations.
  \label{FigRCExamples}}
\end{figure*}
\end{center}

\section{Observed Rotation Curves}
\label{SecObsRotCur}

Our sample of galaxy rotation curves is compiled from several
literature sources. We describe the sources of our compilation in
detail in the Appendix, and list some key properties of the individual
rotation curves that we use in our analysis in
Table~\ref{TabObsGxRotCur}. Our compilation retains only recent
data sets (i.e., published after 2001), and favours, where possible, data sets based on
two-dimensional velocity fields, such as the integral-field optical
data sets of \citet{KuziodeNaray2008}; or the high-resolution ${\rm H\,I}$
datacubes from \citet{Oh2015}. For more massive galaxies, we supplement our compilation with the long-slit rotation curves of \citet{Reyes2011}\footnote{Full rotation curves were not available, we use instead the published parameters of fits to their rotation curves.} and \citet{Kauffmann2015}. Although our
data set is by no means complete, it contains $304$ rotation curves \citep[and an additional $189$ from][]{Reyes2011} spanning a
wide range of $V_{\rm max}$, from $21$ to $350$~km~s$^{-1}$. It also contains
the majority of the dwarf galaxies that have been used in the
literature to illustrate the ``cusp vs core'' problem.

We assume throughout our analysis that the published rotation curves
(which have been corrected, in most cases, for inclination effects,
asymmetric drift, and non circular motions) provide a fair
approximation to the circular velocity profiles of galaxies in our
sample and hereafter refer to both as ``$V_{\rm circ}$''. We note that this is a strong assumption which may fail for a number of
dwarf systems where the observed galaxy is clearly highly irregular.

\section{Observed vs Simulated Rotation Curves}
\label{SecCompRotCur}

\subsection{The similarity of simulated circular velocity curves}

The general properties of the rotation curves of simulated galaxies
shown in Fig.~\ref{FigSimRotCurLog} are in reasonable agreement with
those of observed galaxies, thus extending the agreement between
{\small EAGLE} and observations noted by \citet[][e.g. their fig.~6]{Schaller2014} for $V_{\rm
  max} > 100$~km~s$^{-1}$ to lower mass systems. Circular velocities tend to
rise sharply and stay flat in massive galaxies, but to rise more
slowly in dwarf systems, where baryons play a less important role.

The agreement is not just qualitative. This may be seen in
Fig.~\ref{FigRCExamples}, where we compare directly the rotation
curves of four galaxies of different circular velocity with the
simulation results. The comparison is made with simulated systems
whose maximum circular velocity matches, within $10$ per cent, that of
the observed galaxy, without any rescaling. The match in $V_{\rm max}$
ensures as well that the baryonic masses of simulated and observed
galaxies are comparable, since the simulated systems satisfy the
observed baryonic Tully-Fisher relation (Sales et al 2015, in
preparation).

The excellent agreement shown in Fig. ~\ref{FigRCExamples} is meant to
illustrate a more general point: the rotation curves of {\it many}
galaxies, dwarfs included, are actually consistent with $\Lambda$CDM
predictions. This is important to emphasize, since it is often thought
that $\Lambda$CDM rotation curves are in conflict with data for {\it all} or
{\it a majority} of galaxies, especially dwarfs. 

\subsection{The diversity of observed rotation curves}

\begin{center}
\begin{figure*}
{\leavevmode \includegraphics[width=2.\columnwidth]{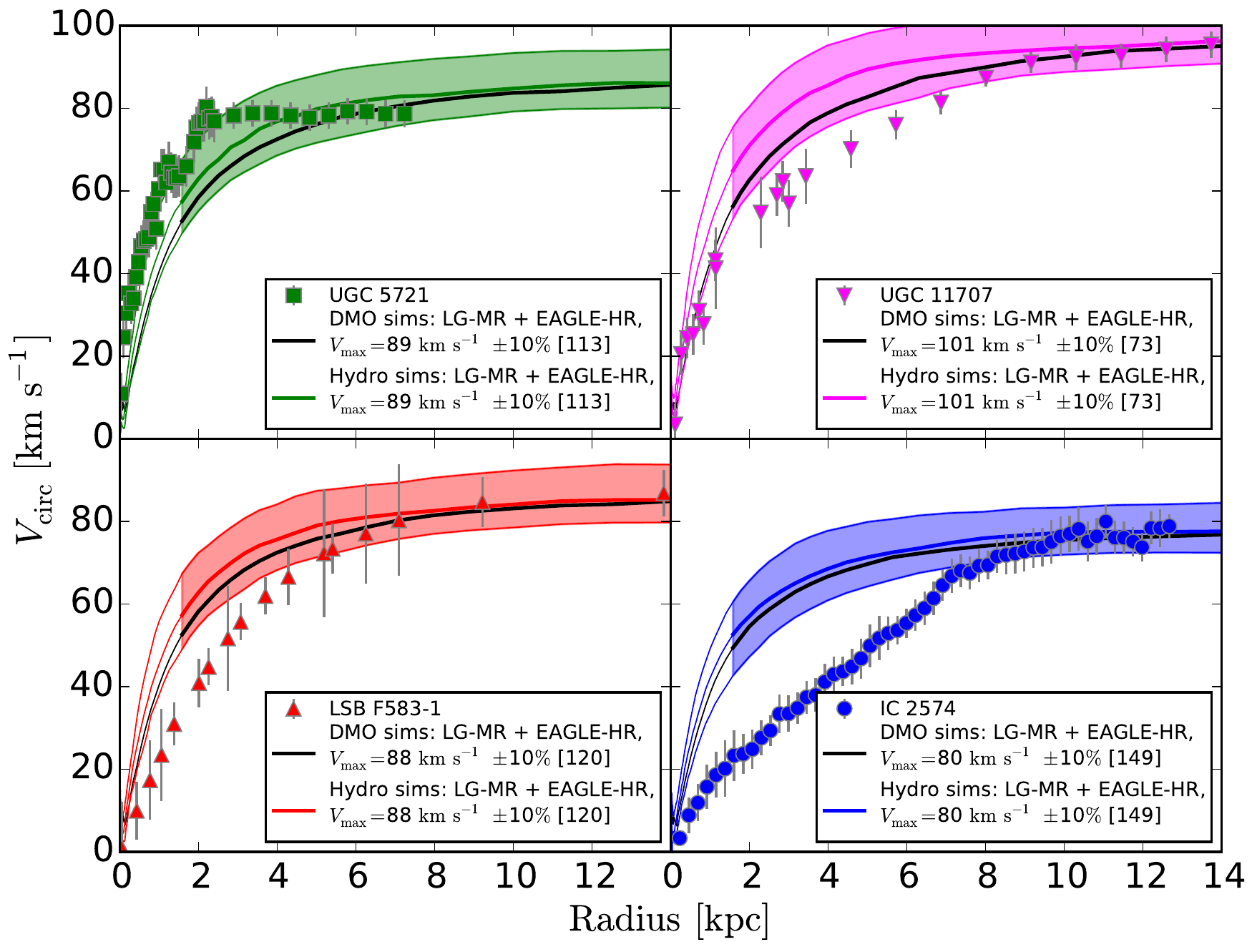}}
\caption{Rotation curves of four dwarf irregular galaxies of
  approximately the same maximum rotation speed ($\sim 80$--$100$~km~s$^{-1}$) and galaxy mass, chosen to illustrate the diversity of rotation curve shape at {\it given} $V_{\rm max}$. As in
  previous figures, coloured solid curves and shaded areas correspond
  to the median (and $10^{\rm th}$--$90^{\rm th}$ percentile) circular velocity curve of
  simulated galaxies matching (within $10$ per cent) the maximum circular
  velocity of each galaxy. Note that the observed rotation curves
  exhibit a much wider diversity than seen in the {\small EAGLE} and {\small LOCAL GROUPS} simulations,
  from galaxies like UGC~5721, which are consistent with our simulations,
  to galaxies like IC~2574, which show a much more slowly rising rotation curve
  compared with simulations, either hydrodynamical (coloured lines) or dark matter-only (black lines).
\label{FigDivRotCur}}
\end{figure*}
\end{center}

Actually, the main difference between simulated and observed rotation
curves is the great {\it diversity} of the latter (especially for
dwarfs), which is unexpected according to our results. We illustrate
this in Fig.~\ref{FigDivRotCur}, where the rotation curves of four
different dwarf galaxies of similar maximum circular velocity are
compared with simulated galaxies of matching $V_{\rm max}$. 

The four galaxies in this figure have been selected to illustrate the
large diversity of rotation curve shapes at fixed $V_{\rm
  max}$. According to the baryonic Tully-Fisher relation
\citep[][]{McGaugh2012}, these four galaxies have similar total
baryonic masses, so the differences in rotation curve shape must be
due to either systematic variations in the spatial distribution of the
baryons, or to varying amounts of dark matter. 

The baryon distribution
is at least partly responsible, since it is well documented that
high-surface brightness galaxies have more steeply rising rotation
curves than low-surface brightness systems \citep[see, e.g.,][and
references therein]{McGaugh1998,Swaters2009}. Quantitatively, however,
the differences cannot be fully ascribed to baryons (see below), so the diversity
seen in Fig.~\ref{FigDivRotCur} reflects large systematic
variations in the inner dark matter content as well.

\subsection{The challenge to $\Lambda$CDM}

The comparison between observed and simulated rotation curves thus
highlights two challenges to $\Lambda$CDM. One is to understand the
origin of the {\it diversity} at fixed $V_{\rm max}$, especially in
dwarf galaxies, which tend to be dark matter dominated.  These are all
galaxies that form in similar halos, have approximately the same
baryonic mass, and similar morphologies. Some diversity induced by
differences in the distribution of the baryonic component is
expected, but clearly the observed diversity is much greater than in
our simulations.

The second, and more worrying, concern is the {\it inner mass deficit}
that some of these galaxies seem to exhibit relative to the $\Lambda$CDM simulation
predictions. Indeed, except for UGC~5721, all of the galaxies shown in
Fig.~\ref{FigDivRotCur} have {\it less} mass in the inner $8$ kpc than
expected not only from our hydro simulations (shaded coloured regions)
but also from a $\Lambda$CDM halo {\it alone} (solid black lines). Systems like
UGC~11707 seem marginally consistent, and could perhaps be interpreted
as outliers, but cases like IC~2574, or LSB~F583-1 are too extreme to
be accommodated by our model without significant change.

The mass deficit we highlight here has been noted before in the
context of the ``cusp vs core'' debate \citep[see, e.g.,][and
references therein]{McGaugh2007}. Indeed, if constant density
``cores'' were imposed on the dark matter it would be relatively
straightforward to reproduce the data shown in
Fig.~\ref{FigDivRotCur}. Such cores, however, would need to vary
from galaxy to galaxy, {\em even at fixed halo mass and galaxy mass}. Indeed, a core at least as large as $\sim 5$ kpc would be needed
to explain the fact that the rotation curve of IC~2574 rises linearly
out to $\sim 8$ kpc, but ought to be much smaller in LSB~F583-1 and even
smaller, if at all present, in UGC~5721.

\subsection{The challenge to baryon-induced core formation}

The diversity of observed rotation curves presents a challenge not
only to our simulations, but also to the baryon-induced ``core''
creation mechanism: why would baryons carve out cores so different in
galaxies that are so similar in terms of morphology, halo mass, and galaxy mass? Further,
we would expect the dark matter to be most affected in systems where
baryons play a more important role in the potential, such as
high-surface brightness galaxies, whereas observations seem to suggest
the opposite trend.

A second challenge concerns the magnitude of the effect needed to
create a core as large as that inferred, for example, for
IC~2574. Published simulations where baryon effects create cores tend
to have overall a modest effect on the total inner mass profile of the
galaxy. One example is provided in Fig.~\ref{FigDG1}; although baryons
have carved a $\sim 1$ kpc core in the dark matter halo in the
simulated galaxy DG1, the total inner mass profile is actually quite
similar to what is expected for galaxies of that circular
velocity in our simulations (green-shaded region), which do not
produce cores. This is because, to first order, the baryons that
displace the dark matter to create a core take its place, leading to
modest net changes in the total mass profile. 

In other words, ``flattening the dark matter cusp'' is not enough to
explain galaxies like IC~2574. A {\it net removal} of large amounts of
mass from the inner regions is needed to reconcile such galaxies with
$\Lambda$CDM, at least if we equate the measured rotation curve with the circular velocity curve. In the case of IC~2574, at least $\sim 2.5\times 10^9\, {\rm M}_\odot$
seem to have been expelled from the inner $\sim 5$ kpc;
more than the total baryonic mass of the galaxy. It seems unlikely
that baryon-induced fluctuations can cause an effect this large.

\begin{figure*}
\leavevmode \includegraphics[width=2.\columnwidth]{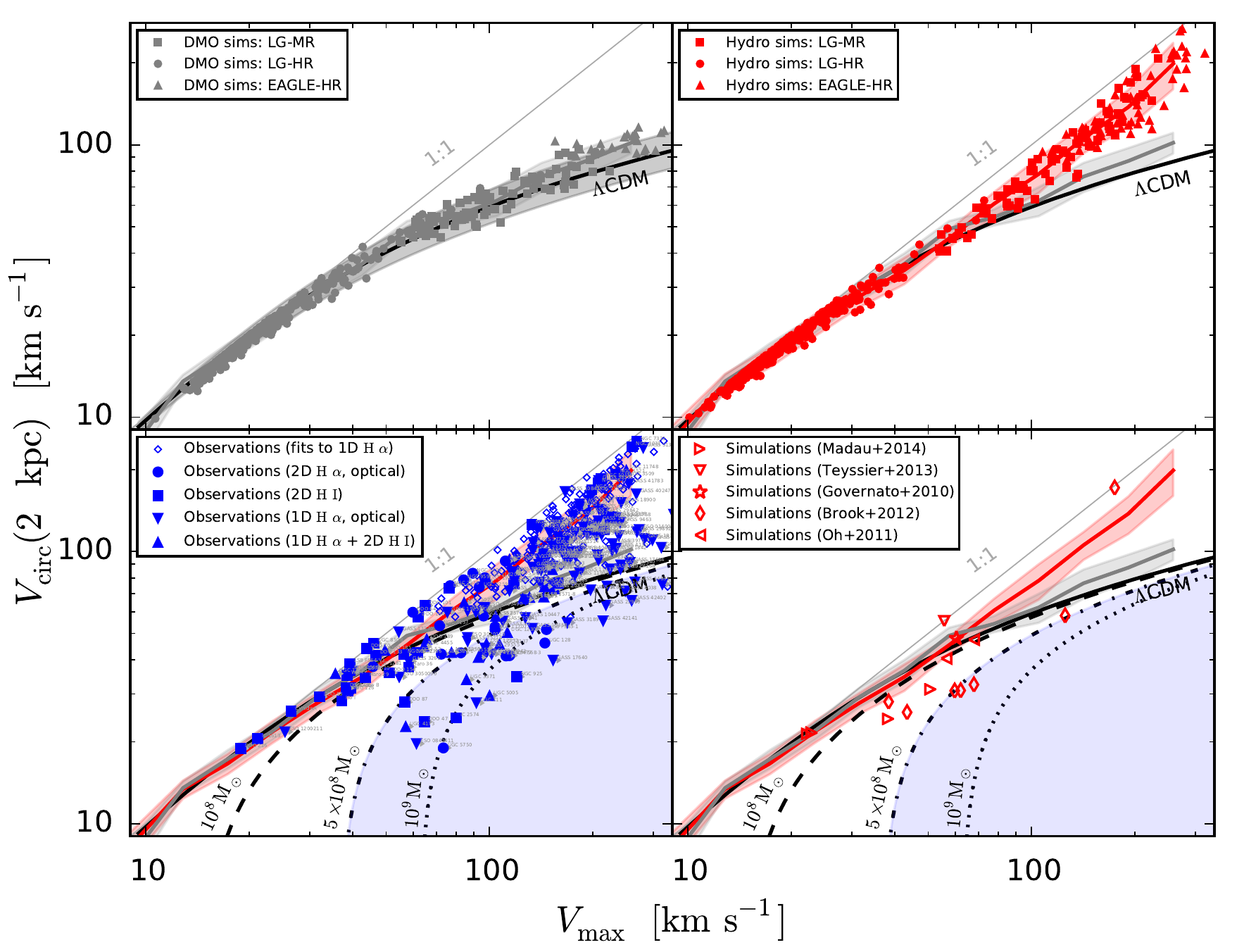}
\caption{Circular velocity at $r=2$~kpc vs the maximum circular
  velocity, $V_{\rm max}$, for observed and simulated galaxies. For
  observed galaxies we use the maximum rotation speed as an
  estimate of $V_{\rm max}$, and the rotation speed measured at $2$~kpc for $V_{\rm circ}(2$~kpc$)$. We show only simulated systems for which the convergence
  radius is less than $2$~kpc, and observed galaxies for which the nominal
  angular resolution of the data is better than the angle subtended by
  $2$~kpc at the galaxy's distance. {\it Top-left:} Results for dark
  matter-only simulations (grey points), together with the
  correlation expected for NFW haloes of average concentration (solid
  black line). The thick gray line traces the mean $V_{\rm circ}(2$~kpc$)$ as a function of
  $V_{\rm max}$, whereas the shaded areas show the standard deviation.
  {\it Top-right:} As the top-left panel, but for simulated galaxies in the
  {\small LOCAL GROUPS} and {\small EAGLE} cosmological hydrodynamical
  simulations (red symbols). See the legend for details about each symbol type. The grey line and grey shaded region repeat the DMO correlation in the top-left
  panel, the red line and shaded region are analogous for the hydrodynamical simulations.
  {\it Bottom-left:} Observed galaxies (small text labels identify individual objects). The different symbols show the different tracers observed (${\rm H\,I}$, ${\rm H}$~$\alpha$, other features in the optical) and whether the observations are in 1 dimension (1D, e.g. long slit spectroscopy) or 2 dimensions (2D, e.g. radio interferometry, integral field spectroscopy). Solid
  lines and shaded regions are as in the
  top right panel. Note the large variation in $V_{\rm circ}(2$~kpc$)$ at fixed $V_{\rm max}$
  compared with the simulation results. The dotted, dashed and dot-dashed lines
  indicate the changes in $V_{\rm circ}(2$~kpc$)$ induced by removing a fixed amount of
  mass from the inner $2$~kpc of $\Lambda$CDM haloes, as labelled. The
  blue-shaded region highlights systems with an inner $2$ kpc mass deficit exceeding 
  $5\times 10^8 M_\odot$.
  {\it Bottom-right:} Results
  of recent simulations that report the formation of cores in the dark
  matter profiles of $\Lambda$CDM haloes. These cores lead to a slight reduction in the
  value of $V_{\rm circ}(2$~kpc$)$ relative to those in our simulations, but the changes
  are insufficient to explain the full range of values spanned by the
  observational data. The dotted lines and dashed lines are as in the
  bottom-left panel, for ease of comparison.
  \label{FigV2Vmax}}
\end{figure*}

\subsection{The challenge to alternative dark matter models}

Finally, we note that the diversity of rotation curves illustrated in
Fig.~\ref{FigDivRotCur} disfavours solutions that rely solely on modifying
the physical nature of the dark matter. Cores can indeed be produced
if the dark matter is SIDM or WDM but, in this case, we would expect
{\it all} galaxies to have cores and, in particular, galaxies of
similar mass or velocity to have cores of similar size. Available
simulation data are sparse but suggest that the scatter in structural
properties at fixed halo mass is no larger for alternative dark matter
models than for $\Lambda$CDM \citep[e.g.][for SIDM and WDM respectively]{Rocha2013,Lovell2014}. This is in
disagreement with rotation curve data and suggests that a mechanism
unrelated to the nature of the dark matter must be invoked to explain
the rotation curve shapes.

\subsection{The ``inner mass deficit'' problem}

The prevalence of the ``inner mass deficit'' problem discussed above
may be characterized by comparing the inner circular velocities of observed
galaxies with those of $\Lambda$CDM galaxies of matching $V_{\rm max}$. We
show this in Fig.~\ref{FigV2Vmax}, where we use our $\Lambda$CDM simulations,
as well as the compiled rotation curve data, to plot the circular
velocity at $2$~kpc against the maximum measured rotation speed,
$V_{\rm max}$. Where data do not exist at exactly $2$~kpc, we interpolate linearly
between nearby data points. We choose a fixed physical radius of $2$~kpc to
characterize the inner mass profile because it is the minimum radius
that is well resolved in all of our simulations for systems in the
mass range of interest here. It is also a radius that is well resolved
in all observed galaxies included in our compilation.

The grey symbols in the top left panel of Fig.~\ref{FigV2Vmax} show
the results of our DMO simulations. The tight correlation between
these quantities in the DMO case is a direct consequence of the nearly
self-similar nature of $\Lambda$CDM haloes: once the cosmological parameters
are specified, the circular velocity at $2$~kpc may be used to predict
$V_{\rm max}$, and vice versa. Variations in environment, shape and
formation history result in some scatter, but overall this is quite
small. For given $V_{\rm max}$, the circular velocity at $2$~kpc has
a standard deviation of only $\sim0.1$~dex.  Our results are in good agreement with
earlier DMO simulation work.  The solid black line (and shaded region)
in the figure indicates the expected correlation (plus 1-$\sigma$
scatter) for NFW haloes with the mass-concentration relation
corresponding to the cosmological parameters adopted in our
simulations \citep{Ludlow2014}. Note that the simulated data approach
the $1$:$1$ line for $V_{\rm max}<30$ km/s: this is because those
halos are intrinsically small; the radius where circular
velocity profiles peak decreases steadily with decreasing circular
velocity, from $4.6$~kpc to $1.9$~kpc when $V_{\rm max}$ decreases from
$30$ to $15$ km/s.

The inclusion of baryons modifies these correlations, as shown by the
red symbols in the top-right panel of Fig.~\ref{FigV2Vmax}, which show
results for our hydrodynamical simulations. The main result of
including baryons is to shift the expected correlation toward higher
values of $V_{\rm circ}(2$ kpc$)$ for galaxies with $V_{\rm max} \gtrsim 60$~km~s$^{-1}$. This is not surprising: the
assembly of the luminous galaxy adds mass to the central few
kiloparsecs and raises the circular velocity there. A tight relation
between $V_{\rm max}$ and $V_{\rm circ}(2$~kpc$)$ remains, however:
the scatter increases only slightly, to at most $\sim 0.15$~dex
(standard deviation).

Observed galaxies are shown in the bottom-left panel of
Fig.~\ref{FigV2Vmax}. The diversity of rotation curves alluded to
above is clearly seen here. At $V_{\rm max}\sim 70$~km~s$^{-1}$, for
example, the rotation speed at $2$~kpc of observed galaxies spans more
than a factor of $\sim 4$, or about a factor of $\sim 16$ in enclosed
mass. Some of those galaxies, like DDO~168 have rotation speeds at
$2$~kpc comparable to the maximum ($V_{\rm max}\sim 62$~km~s$^{-1}$,
$V_{\rm circ}(2$~kpc$) \sim 58$~km~s$^{-1}$), which indicates an
enclosed mass of $\sim 2.3\times10^{9}\,{\rm M}_\odot$, or about twice
as much as the {\it total} baryonic mass of the galaxy, according to
the baryonic Tully-Fisher relation; $M_{\rm bar}/{\rm M}_\odot = 102.3
\, (V_{\rm max}/$~km~s$^{-1})^{3.82}$ \citep[][]{McGaugh2012}. At the
other extreme, galaxies like UGC~5750 ($V_{\rm max}\gtrsim
73$~km~s$^{-1}$) \footnote{A rightward arrow is used in the bottom left panel
of Fig.~\ref{FigV2Vmax} to indicate cases where the rotation curve is
still rising at the outermost radius measured -- the maximum observed
rotation speed may therefore underestimate $V_{\rm max}$.} have
rotation speeds at $2$~kpc of just $\sim 20$~km~s$^{-1}$,
corresponding to an enclosed mass of only $\sim 2\times 10^{8}\,{\rm
  M}_\odot$, or just about $10$ per cent of its total baryonic mass.

Within their diversity, many observed galaxies actually have rotation curves
that agree with $\Lambda$CDM, and fall well within the region of parameter
space expected from our simulations (shown by the red-shaded area in
this panel). Others do not. Galaxies {\it below} the solid gray line
(which indicates the average DMO results in all panels) have less mass
within $2$~kpc than expected from a DMO simulation: if
rotation velocities faithfully represent the circular velocity at this
radius, then some of the central mass must have been displaced.

The dotted, dot-dashed, and dashed lines in the bottom panels of
Fig.~\ref{FigV2Vmax} quantify this effect. They indicate the result of
removing a fixed amount of mass, as labelled, from the inner $2$~kpc
of NFW halos on the $\Lambda$CDM correlation (solid black
line). Galaxies in the light blue-shaded area below the dot-dashed
curve, in particular, have a mass deficit in the inner $2$ kpc
of more than $5\times 10^8 M_\odot$ compared with a typical
$\Lambda$CDM halo. For DDO~87 ($V_{\rm max}\gtrsim 57$~km~s$^{-1}$,
$V_{\rm circ}(2$~kpc$)\sim 28$~km~s$^{-1}$), for example, a galaxy that
falls close to the dot-dashed line, this implies a total mass deficit in
the inner $2$ kpc comparable to its total baryonic mass, as estimated
from the baryonic Tully-Fisher relation. Indeed, several galaxies in
our compilation have apparently expelled from the inner $2$~kpc a mass
comparable to or larger than their total baryonic mass.

Published simulations that report baryon-induced ``cores'' seem unable
to match these results. We show this in the bottom-right panel of
Fig.~\ref{FigV2Vmax}, where the different open symbols show the
results of simulations for which the formation of a ``core'' in the
dark matter has been reported. A few trends seem clear. Although
``core'' formation alleviates the problem in some cases by bringing
down the velocities at $2$~kpc, the effect is generally small; indeed,
no ``cored'' galaxies lie well inside the blue shaded area that
characterizes systems with a mass deficit larger than $5\times 10^8
M_\odot$ in the inner $2$ kpc .

Further, core creation -- or inner mass removal -- seems ineffective
in galaxies with $V_{\rm max}>100$~km s$^{-1}$.  Observations, on the
other hand, show sizeable ``inner mass deficits'' even in galaxies
with maximum rotation velocities well above $100$~km~s$^{-1}$. This
may be a problem for ``baryon-induced'' core formation, since it has
been argued that the potential well might be too deep\footnote{We note
  that \citet{Maccio2012b} report the creation of cores in more
  massive systems, but since these authors do not show the circular
  velocity profiles, we have been unable to add their results to our
  plot.} in such systems for baryons to create a sizeable core
\citep[][]{Brook2012}. We note, however, that this conclusion is based
only on two systems (see bottom-right panel of Fig.~\ref{FigV2Vmax}),
and that none of those simulations include AGN feedback. It remains to
be seen whether further simulation work will be able to produce inner
mass deficits as large as observed in some of these massive galaxies.

With these caveats in mind, we conclude that none of the mechanisms
proposed so far to explain the apparent presence of cores in dwarf
galaxies has been able to fully account for their inner mass deficits
and for the observed diversity of their rotation curves.

\section{Summary and conclusions}
\label{SecConc}

We have used circular velocity curves from recent cosmological
hydrodynamical simulations of galaxy formation in a $\Lambda$CDM
universe, taken from the {\small EAGLE} and {\small LOCAL GROUPS} projects, to investigate the
rotation curves of galaxies and reassess the ``cusp vs core''
controversy.

The circular velocity curves of simulated galaxies vary systematically
as a function of their maximum circular velocity ($V_{\rm max}$), but show little
variation for given $V_{\rm max}$.  Observed rotation curves, on the
other hand, show great diversity, even at fixed $V_{\rm max}$, 
especially for dwarf galaxies. At any given maximum rotation speed,
some have shapes that are consistent with the simulation predictions,
others do not. Deviant galaxies typically have much lower circular
velocities in the inner regions than expected in $\Lambda$CDM from the
dark matter halo {\it alone}. This apparent inner mass deficit varies
from galaxy to galaxy, even at fixed galaxy mass, and can exceed,
within $\sim 2$ kpc, the total baryonic mass of a galaxy.

Although this inner mass deficit may also be interpreted as evidence
for a ``core'' in the dark matter profile, we argue that
characterizing the problem as an inner mass deficit is more
robust, since it allows simulations to be compared directly with data
without relying on uncertain decomposition of the dark matter and
baryonic contributions to the central mass profile, or estimating
dark matter density slopes in the innermost regions, where
uncertainties in observations and simulations are largest.

Models that attempt to reconcile rotation curves with $\Lambda$CDM by carving
``cores'' in the dark matter through baryon-induced gravitational
fluctuations offer no natural explanation for the large dispersion in
the values of the observed mass deficit.  Nor do they seem able, at least
according to published simulations, to account quantitatively for the largest
mass deficits observed.

The diversity of observed rotation curves is also unexpected in
alternative dark matter scenarios, substantially diminishing their
appeal. This is because modifying the nature of dark matter may
produce cores in dark haloes, but such cores would all be of similar
size at given mass scale, unlike what is inferred from rotation
curves.

Finally, it may be that dynamical inferences from available kinematic
data need to be reevaluated. Many of the galaxies that show the
largest mass deficits (or the strongest evidence for a ``core'')
appear highly irregular. Complexities such as non-circular and random
motions, instrumental smearing and sampling effects, and/or departures
from axisymmetry and coplanarity can substantially complicate the
reconstruction of circular velocity profiles from the observed
kinematics. The magnitude of the effect needed to reconcile
$\Lambda$CDM with the data shown in Fig.~\ref{FigV2Vmax} seem too
large, however, to be due wholly to such uncertainties.

We conclude that present rotation curve data support neither a revision
of the nature of dark matter, nor current models for ``core
formation'' in galaxies formed in $\Lambda$CDM universe. The mystery
of the inner rotation curves of galaxies thus remains unsolved.

\section*{Acknowledgements}\label{sec-awknowledgements}

We thank S.-H.~Oh, E.~de~Blok, S.~McGaugh, C.~Brook, J.~Adams, S.~Moran and G.~Kauffmann for data
contributions. This work was supported in part by the
Science and Technology Facilities Council (grant number ST/F001166/1);
European Research Council (grant numbers GA 267291 ``Cosmiway'' and GA
278594 ``Gas Around Galaxies''). This research was supported in part by the
National Science Foundation under Grant No. NSF PHY11-25915, the Interuniversity Attraction Poles Programme initiated by the Belgian Science Policy Office ([AP P7/08 CHARM]) and by the
hospitality of the Kavli Institute for Theoretical Physics at the
University of California, Santa Barbara. RAC is a Royal Society University Research Fellow.

\bibliography{paper}

\appendix
\section{Observed Rotation Curve Compilation}
\label{SecApp}

Our sample of galaxy rotation curves is drawn from several sources. We describe each set of observations in \S\S~\ref{firstsubsec}--\ref{lastsubsec}; the key properties of the individual rotation curves are summarized in Table~\ref{TabObsGxRotCur}.

\subsection{\citet{Kauffmann2015}}\label{firstsubsec}

This publication presents 187 rotation curves for galaxies in the GALEX Arecibo Sloan Survey (GASS) with maximum rotation velocities of 90--350~km~s$^{-1}$. Long slit spectra were measured using the blue-channel spectrograph on the 6.5-m Multi Mirror Telescope and the Dual Imaging Spectrograph on the 3.5-m Apache Point Observatory telescope, with angular resolutions of $1.25$ arcsec and $1.5$ arcsec respectively. Though long slit spectroscopy offers excellent angular resolution, the main drawback is that the velocity field is measured only along one direction through the galaxy, making possible errors due to, for example, non-circular motions more difficult to quantify. For each spectrum, an attempt was made to produce two rotation curves: one derived from ${\rm H}$~$\alpha$ emission, and a second from fitting stellar absorption templates. The stellar absorption based rotation curves are typically in good agreement with the ${\rm H}$~$\alpha$ curves, but of slightly better quality, so we use these where possible. Of the 187 rotation curves in the sample, we retain 106 in our compilation, the rest being of insufficient quality, resolution or extent for use in our analysis. Of these, 99 are stellar absorption based rotation curves (of which 52 also have good ${\rm H}$~$\alpha$ rotation curves, which we discard) and 7 are ${\rm H}$~$\alpha$ based. For brevity, we omit all discarded rotation curves from this source in Table~\ref{TabObsGxRotCur}.

\subsection{\citet{Oh2015}}

This publication presents 26 rotation curves for galaxies in the Local Irregulars That Trace Luminosity Extremes, The ${\rm H\,I}$ Nearby Galaxy Survey (LITTLE THINGS) sample. This observing program is closely related to the THINGS survey (see \S\ref{thingssubsec}). The maximum measured rotation velocities of these galaxies range from about 20 to 120~km~s$^{-1}$. Observations were taken using the NRAO Very Large Array. The angular resolution of $6$ arcsec is a factor of 2 better than that of the THINGS survey. Radio interferometry leads naturally to 2-dimensional
velocity maps. ${\rm H\,I}$ observations are traditionally better than
their optical (usually ${\rm H}$~$\alpha$) counterparts at mapping the velocity field in
the outskirts of galaxies at the cost of poorer spatial resolution
throughout. The LITTLE THINGS sample offers a good compromise, with
sufficient resolution to probe the inner kiloparsec. The rotation
curves were constructed from the velocity fields using a tilted-ring
model, with asymmetric drift corrections applied as needed. We retain all 26 galaxies in our sample.

\subsection{\citet{Adams2014}}

This publication presents a sample of 7 rotation curves of galaxies with maximum rotation velocities of about 100~km~s$^{-1}$. The velocity fields were measured with the VIRUS-W integral field spectrograph on the 2.7-m Harlan J. Smith telescope at McDonald Observatory.  The use of an integral field unit (IFU) allows measurement of the velocity field in 2 dimensions, while maintaining the high spatial resolution typical of optical rotation curves: the fibres feeding the spectrograph have diameters of $3.1$ arcsec. The velocity field is constructed by simultaneous mapping of the ${\rm H}$~$\beta$ line and two ${\rm O}\,{\rm III}$ lines. The rotation curves were constructed from the velocity fields using a tilted-ring model. We retain all 7 galaxies in our sample.

\subsection{\citet{deBlok2008} and \citet{Oh2011}}\label{thingssubsec}

These two publications are part of the THINGS project. The 19 rotation curves presented in \citet{deBlok2008} are those that are most straightforwardly derived,
while seven of those requiring more careful analysis were presented in
\citet{Oh2011} \citep[in some cases reanalysing galaxies from ][]{deBlok2008}. Maximum measured rotation velocities are 30--300~km~s$^{-1}$. The survey was carried out with the NRAO Very Large
Array in B, C and D configurations. The angular resolution of $12$ arcsec
is substantially better than that of most earlier ${\rm H}\,{\rm I}$ rotation curve
measurements. The rotation
curves were constructed from the velocity fields using a tilted-ring
model. Four galaxies from \citet{Oh2011} -- Ho~I, Ho~II, M81dwB and DDO~53 -- required substantial
asymmetric drift corrections. We retain 15 of the 19 galaxies from
\citet{deBlok2008}, discarding three that were reanalyzed by
\citet{Oh2011} and NGC~4826, which appears to have a counter-rotating
disk component that complicates the interpretation. We discard Ho II (also named DDO~50), DDO~53, DDO~154 and NGC~2366 from the \citet{Oh2011} sample in favour of the higher angular resolution observations of \citet{Oh2015}.

\subsection{\citet{Reyes2011}}

This publication presents a collection of 189 rotation curves with maximum rotation velocities of 100--400~km~s$^{-1}$. All are long slit ${\rm H}$~$\alpha$ measurements, variously measured using the TWIN spectrograph on the 3.5-m telescope at Calar Alto Observatory ($\sim1.5$ arcsec resolution), the CCDS spectrograph on the 2.4-m Hiltner telescope at the MDM Observatory ($\sim2$ arcsec resolution), and the Dual Imaging Spectrograph on the 3.5-m telescope at Apache Point Observatory ($\sim1.5$ arcsec resolution). We were not able to obtain the full rotation curve dataset, but use instead the parameters of an arctangent model fit \citep{Courteau1997} to the rotation curves to estimate the rotation velocity at $2$~kpc and at maximum (the asymptotic value of the model). We retain all 189 rotation curves in our sample, but use an open symbol type in Fig.~\ref{FigV2Vmax} to visually separate these data from the other sources in our compilation for which we have full rotation curves.

\subsection{\citet{KuziodeNaray2008}}

This publication presents rotation curves for a selection of 17
galaxies with maximum rotation velocities of 50--100~km~s$^{-1}$. These were measured with the DensePak IFU
on the 3.5-m WIYN telescope at the Kitt Peak National Observatory. The
instrument has a pixel size of $3$ arcsec, and multiple slightly offset
exposures of the same region were taken to improve the resolution of
the rotation curves to $\sim2$ arcsec. The rotation
curves were constructed from the velocity fields using a tilted-ring
model. No correction for asymmetric drift was applied as the correction
is expected to be $<2$~km~s$^{-1}$ in all cases. We retain 15 galaxies in our sample, discarding UGC~4325 (alternate name NGC~2552) and NGC~959 in favour of more recent observations.

\subsection{\citet{deBlok2004}}

This publication presents a single, very high resolution rotation
curve. Discounting the satellites of the Milky Way, NGC~6822 is the
nearest late-type dwarf galaxy to us, at a distance of less than 500
kpc. This allows for a high spatial resolution ($\sim 20$ pc) ${\rm
  H\,I}$ map even with the modest $8$ arcsec angular resolution of the
Australia Telescope Compact Array.

\subsection{\citet{Swaters2003}}

This publication presents 15 rotation curves with maximum measured
rotation velocities of about 100--150~km~s$^{-1}$. The double spectrograph on
the 200-inch Hale telescope at Mount Palomar Observatory was used to
obtain long-slit ${\rm H}$~$\alpha$ spectra along the major axis of each
galaxy. The spectra have a nominal resolution of $0.5$ arcsec but, in
practice, are limited by seeing of $\sim 1$--$2$ arcsec. The galaxies
targetted also had existing ${\rm H\,I}$ maps, which were used to extend
the ${\rm H}$~$\alpha$ curves to larger radii. We retain 4 galaxies in our sample,
discarding UGC~731 that has a counter-rotating component, and UGC~8490, UGC~5721, UGC~4499, UGC~4325, UGC~2259, UGC~11861, UGC~11707, UGC~11557, LSB~F568-3 and LSB~F563-V2 in favour of more recent observations of the same galaxies.

\subsection{\citet{deBlok2002}}

This publication presents 24 rotation curves of galaxies with maximum
measured rotation velocities of about 50--100~km~s$^{-1}$. Long slit ${\rm H}$~$\alpha$ spectra were measured using the Carelec spectrograph on
the 192-cm telescope at the Observatoire de Haute Provence. The
angular resolution is seeing-limited to $\sim2$ arcsec. These data are
supplemented with lower resolution ${\rm H\,I}$ velocity maps for all
but two of the galaxies, allowing the extension of the rotation curves
to larger radii. We exclude UGC~5750, UGC~4325, UGC~1281, UGC~10310, NGC~4395, NGC~3274, NGC~2366, LSB~F563-1 and DDO~64 from our sample in favour
of more recent observations, retaining 15 galaxies.

\subsection{\citet{deBlok2001}}\label{lastsubsec}

This paper presents 26 rotation curves of galaxies with maximum
measured rotation velocities between 40 and 200~km~s$^{-1}$. The 4-m
Kitt Peak and 100-in Las Campanas telescopes were used to obtain long
slit ${\rm H}$~$\alpha$ spectra. The resolution is seeing-limited to
$\sim1.5$ arcsec. Where available, ${\rm H\,I}$ velocity maps supplement the
${\rm H}$~$\alpha$ data, extending the rotation curves to larger
radii. We exclude UGC~5750, UGC~11557, LSB~F583-4, LSB~F583-1, LSB~F568-3 and LSB~F563-1 from our sample in favour of more recent
observations, retaining 20 rotation curves.

\newpage

\onecolumn

\LTcapwidth=\textwidth

\begin{longtable}{llrrrrrr}

\caption{Basic properties of galaxies in our sample. Galaxy names and distances are those used in the publication cited. The observation type specifies the spectral feature(s) used to construct the rotation curve. Resolutions assume the distance in column 3, based on angular resolution as quoted by the cited publication. $V_{\rm circ}(2$~kpc$)$ and $V_{\rm max}$ are the quantities plotted in Fig.~\ref{FigV2Vmax}. Explanation of notes: (1) stellar absorption-derived rotation curve preferred to ${\rm H}$~$\alpha$-derived rotation curve (see \S\ref{firstsubsec}); (2) galaxy excluded from our sample in favour of a more recent observation; (3) excluded due to a counter-rotating component in the rotation curve.}\label{TabObsGxRotCur}\vspace{.3cm}\\

 & & Distance & Observation & Observation & $V_{\rm circ}(2 {\rm kpc})$ & $V_{\rm max}$ & \\ 
\multicolumn{1}{c}{Galaxy} & \multicolumn{1}{c}{Reference} & (Mpc) & type & resolution (kpc) & (km s$^{-1}$) & (km s$^{-1}$) & Notes\\
\hline
\endfirsthead

 & & Distance & Observation & Observation & $V_{\rm circ}(2 {\rm kpc})$ & $V_{\rm max}$ & \\ 
\multicolumn{1}{c}{Galaxy} & \multicolumn{1}{c}{Reference} & (Mpc) & type & resolution (kpc) & (km s$^{-1}$) & (km s$^{-1}$) & Notes\\
\hline
\endhead

\hline \multicolumn{8}{r}{{\emph{Continued on next page.}}} \\
\endfoot

\hline
\endlastfoot

GASS 9891&\cite{Kauffmann2015}&$110.5$&${\rm stellar\,abs.}$&$0.8$&$122.3$&$\geq353.0$&$(1)$\\
GASS 9463&\cite{Kauffmann2015}&$152.5$&${\rm H}\;\alpha$&$1.1$&$127.6$&$\geq239.9$&$-$\\
GASS 8096&\cite{Kauffmann2015}&$147.8$&${\rm stellar\,abs.}$&$0.9$&$122.7$&$\geq187.4$&$(1)$\\
GASS 7286&\cite{Kauffmann2015}&$115.2$&${\rm stellar\,abs.}$&$0.7$&$85.3$&$141.2$&$(1)$\\
GASS 7031&\cite{Kauffmann2015}&$141.8$&${\rm stellar\,abs.}$&$0.9$&$99.7$&$171.2$&$-$\\
GASS 6583&\cite{Kauffmann2015}&$206.9$&${\rm stellar\,abs.}$&$1.2$&$88.3$&$\geq152.9$&$-$\\
GASS 57017&\cite{Kauffmann2015}&$138.3$&${\rm H}\;\alpha$&$1.0$&$96.2$&$\geq178.0$&$-$\\
GASS 56612&\cite{Kauffmann2015}&$124.2$&${\rm stellar\,abs.}$&$0.7$&$128.6$&$\geq156.6$&$-$\\
GASS 52297&\cite{Kauffmann2015}&$140.5$&${\rm stellar\,abs.}$&$0.8$&$84.1$&$\geq116.5$&$-$\\
GASS 51899&\cite{Kauffmann2015}&$165.7$&${\rm stellar\,abs.}$&$1.0$&$93.9$&$139.8$&$(1)$\\
GASS 51416&\cite{Kauffmann2015}&$190.6$&${\rm stellar\,abs.}$&$1.1$&$77.8$&$118.3$&$-$\\
GASS 51351&\cite{Kauffmann2015}&$125.5$&${\rm stellar\,abs.}$&$0.9$&$240.1$&$280.0$&$(1)$\\
GASS 48356&\cite{Kauffmann2015}&$122.9$&${\rm stellar\,abs.}$&$0.9$&$124.4$&$139.0$&$(1)$\\
GASS 47221&\cite{Kauffmann2015}&$136.2$&${\rm stellar\,abs.}$&$0.8$&$77.0$&$\geq140.4$&$(1)$\\
GASS 42402&\cite{Kauffmann2015}&$197.0$&${\rm stellar\,abs.}$&$1.2$&$66.0$&$\geq259.1$&$-$\\
GASS 42141&\cite{Kauffmann2015}&$154.2$&${\rm stellar\,abs.}$&$1.1$&$55.6$&$214.1$&$-$\\
GASS 42140&\cite{Kauffmann2015}&$195.7$&${\rm stellar\,abs.}$&$1.2$&$89.2$&$\geq240.9$&$-$\\
GASS 42025&\cite{Kauffmann2015}&$157.2$&${\rm stellar\,abs.}$&$0.9$&$86.1$&$\geq224.3$&$(1)$\\
GASS 41783&\cite{Kauffmann2015}&$158.9$&${\rm stellar\,abs.}$&$1.1$&$177.4$&$254.3$&$(1)$\\
GASS 4137&\cite{Kauffmann2015}&$190.6$&${\rm stellar\,abs.}$&$1.1$&$66.5$&$\geq127.3$&$-$\\
GASS 41323&\cite{Kauffmann2015}&$188.4$&${\rm stellar\,abs.}$&$1.1$&$41.3$&$96.4$&$-$\\
GASS 4130&\cite{Kauffmann2015}&$191.9$&${\rm stellar\,abs.}$&$1.4$&$102.8$&$233.8$&$-$\\
GASS 4094&\cite{Kauffmann2015}&$118.2$&${\rm stellar\,abs.}$&$0.7$&$116.4$&$193.3$&$-$\\
GASS 4057&\cite{Kauffmann2015}&$170.0$&${\rm stellar\,abs.}$&$1.0$&$138.9$&$\geq192.0$&$-$\\
GASS 4048&\cite{Kauffmann2015}&$177.3$&${\rm stellar\,abs.}$&$1.1$&$100.3$&$\geq182.1$&$(1)$\\
GASS 4040&\cite{Kauffmann2015}&$115.2$&${\rm stellar\,abs.}$&$0.7$&$122.7$&$140.6$&$-$\\
GASS 4038&\cite{Kauffmann2015}&$178.6$&${\rm stellar\,abs.}$&$1.1$&$71.5$&$\geq248.4$&$(1)$\\
GASS 40317&\cite{Kauffmann2015}&$174.7$&${\rm stellar\,abs.}$&$1.0$&$153.8$&$195.4$&$(1)$\\
GASS 40257&\cite{Kauffmann2015}&$168.3$&${\rm stellar\,abs.}$&$1.0$&$75.3$&$\geq150.0$&$-$\\
GASS 40247&\cite{Kauffmann2015}&$167.9$&${\rm stellar\,abs.}$&$1.2$&$162.6$&$267.4$&$-$\\
GASS 3971&\cite{Kauffmann2015}&$182.4$&${\rm stellar\,abs.}$&$1.3$&$106.8$&$228.4$&$(1)$\\
GASS 39595&\cite{Kauffmann2015}&$186.3$&${\rm stellar\,abs.}$&$1.1$&$182.0$&$\geq208.0$&$(1)$\\
GASS 39567&\cite{Kauffmann2015}&$133.6$&${\rm stellar\,abs.}$&$0.8$&$145.4$&$\geq182.9$&$(1)$\\
GASS 38964&\cite{Kauffmann2015}&$137.9$&${\rm stellar\,abs.}$&$1.0$&$110.2$&$297.9$&$(1)$\\
GASS 38758&\cite{Kauffmann2015}&$124.6$&${\rm stellar\,abs.}$&$0.9$&$132.9$&$233.7$&$(1)$\\
GASS 38472&\cite{Kauffmann2015}&$113.1$&${\rm stellar\,abs.}$&$0.7$&$89.7$&$139.1$&$-$\\
GASS 3819&\cite{Kauffmann2015}&$194.0$&${\rm stellar\,abs.}$&$1.2$&$60.6$&$86.3$&$(1)$\\
GASS 3817&\cite{Kauffmann2015}&$192.7$&${\rm stellar\,abs.}$&$1.2$&$71.8$&$\geq114.3$&$(1)$\\
GASS 3777&\cite{Kauffmann2015}&$169.6$&${\rm stellar\,abs.}$&$1.0$&$107.8$&$\geq134.1$&$(1)$\\
GASS 3645&\cite{Kauffmann2015}&$131.5$&${\rm stellar\,abs.}$&$0.8$&$115.2$&$211.9$&$(1)$\\
GASS 3524&\cite{Kauffmann2015}&$162.7$&${\rm stellar\,abs.}$&$1.0$&$114.1$&$237.9$&$(1)$\\
GASS 3524&\cite{Kauffmann2015}&$162.7$&${\rm H}\;\alpha$&$1.0$&$122.4$&$\geq199.9$&$-$\\
GASS 3509&\cite{Kauffmann2015}&$207.3$&${\rm stellar\,abs.}$&$1.2$&$187.4$&$244.8$&$(1)$\\
GASS 3439&\cite{Kauffmann2015}&$165.3$&${\rm stellar\,abs.}$&$1.0$&$113.0$&$158.3$&$-$\\
GASS 3261&\cite{Kauffmann2015}&$160.6$&${\rm stellar\,abs.}$&$1.0$&$39.5$&$\geq58.2$&$-$\\
GASS 3189&\cite{Kauffmann2015}&$164.5$&${\rm H}\;\alpha$&$1.0$&$54.7$&$\geq168.3$&$-$\\
GASS 30811&\cite{Kauffmann2015}&$209.4$&${\rm stellar\,abs.}$&$1.3$&$103.4$&$\geq316.1$&$-$\\
GASS 30479&\cite{Kauffmann2015}&$131.9$&${\rm stellar\,abs.}$&$0.8$&$121.1$&$\geq137.8$&$(1)$\\
GASS 30338&\cite{Kauffmann2015}&$179.0$&${\rm stellar\,abs.}$&$1.1$&$134.1$&$\geq227.4$&$(1)$\\
GASS 29892&\cite{Kauffmann2015}&$156.7$&${\rm stellar\,abs.}$&$0.9$&$118.1$&$268.5$&$(1)$\\
GASS 29842&\cite{Kauffmann2015}&$146.0$&${\rm stellar\,abs.}$&$1.1$&$138.0$&$215.3$&$(1)$\\
GASS 29555&\cite{Kauffmann2015}&$135.3$&${\rm stellar\,abs.}$&$0.8$&$44.2$&$\geq97.1$&$(1)$\\
GASS 27167&\cite{Kauffmann2015}&$162.7$&${\rm stellar\,abs.}$&$1.0$&$134.6$&$175.1$&$(1)$\\
GASS 26822&\cite{Kauffmann2015}&$161.0$&${\rm stellar\,abs.}$&$1.2$&$97.2$&$206.0$&$(1)$\\
GASS 25214&\cite{Kauffmann2015}&$133.2$&${\rm stellar\,abs.}$&$0.8$&$46.8$&$86.4$&$-$\\
GASS 24496&\cite{Kauffmann2015}&$180.3$&${\rm stellar\,abs.}$&$1.1$&$121.2$&$\geq198.2$&$(1)$\\
GASS 24366&\cite{Kauffmann2015}&$176.0$&${\rm stellar\,abs.}$&$1.1$&$123.7$&$\geq191.7$&$-$\\
GASS 24168&\cite{Kauffmann2015}&$111.4$&${\rm stellar\,abs.}$&$0.7$&$86.3$&$\geq240.5$&$(1)$\\
GASS 24094&\cite{Kauffmann2015}&$184.6$&${\rm stellar\,abs.}$&$1.1$&$149.5$&$\geq183.1$&$-$\\
GASS 23450&\cite{Kauffmann2015}&$203.9$&${\rm stellar\,abs.}$&$1.2$&$92.9$&$207.5$&$-$\\
GASS 23315&\cite{Kauffmann2015}&$140.9$&${\rm stellar\,abs.}$&$0.8$&$101.0$&$\geq133.6$&$(1)$\\
GASS 22999&\cite{Kauffmann2015}&$194.9$&${\rm stellar\,abs.}$&$1.2$&$63.5$&$\geq218.1$&$(1)$\\
GASS 21842&\cite{Kauffmann2015}&$192.7$&${\rm stellar\,abs.}$&$1.2$&$100.7$&$227.4$&$-$\\
GASS 20292&\cite{Kauffmann2015}&$128.1$&${\rm stellar\,abs.}$&$0.8$&$93.7$&$166.6$&$-$\\
GASS 20133&\cite{Kauffmann2015}&$209.4$&${\rm stellar\,abs.}$&$1.3$&$73.1$&$\geq151.5$&$-$\\
GASS 20041&\cite{Kauffmann2015}&$132.3$&${\rm stellar\,abs.}$&$0.8$&$55.4$&$\geq109.4$&$(1)$\\
GASS 18900&\cite{Kauffmann2015}&$194.0$&${\rm stellar\,abs.}$&$1.2$&$151.0$&$\geq241.4$&$-$\\
GASS 18335&\cite{Kauffmann2015}&$184.6$&${\rm stellar\,abs.}$&$1.1$&$111.5$&$\geq281.8$&$-$\\
GASS 17684&\cite{Kauffmann2015}&$154.6$&${\rm stellar\,abs.}$&$0.9$&$91.1$&$\geq256.2$&$(1)$\\
GASS 17640&\cite{Kauffmann2015}&$149.5$&${\rm stellar\,abs.}$&$0.9$&$39.8$&$\geq153.1$&$-$\\
GASS 15257&\cite{Kauffmann2015}&$123.3$&${\rm stellar\,abs.}$&$0.7$&$59.8$&$\geq75.4$&$-$\\
GASS 15181&\cite{Kauffmann2015}&$200.4$&${\rm stellar\,abs.}$&$1.2$&$74.5$&$\geq219.8$&$-$\\
GASS 14831&\cite{Kauffmann2015}&$190.2$&${\rm stellar\,abs.}$&$1.1$&$77.2$&$265.2$&$(1)$\\
GASS 14247&\cite{Kauffmann2015}&$141.3$&${\rm stellar\,abs.}$&$1.0$&$75.8$&$\geq318.3$&$-$\\
GASS 14017&\cite{Kauffmann2015}&$169.6$&${\rm stellar\,abs.}$&$1.0$&$129.3$&$\geq207.2$&$-$\\
GASS 12460&\cite{Kauffmann2015}&$211.6$&${\rm stellar\,abs.}$&$1.3$&$136.9$&$\geq343.4$&$(1)$\\
GASS 12069&\cite{Kauffmann2015}&$166.2$&${\rm stellar\,abs.}$&$1.2$&$86.3$&$\geq138.6$&$(1)$\\
GASS 12002&\cite{Kauffmann2015}&$157.2$&${\rm stellar\,abs.}$&$0.9$&$142.7$&$208.3$&$-$\\
GASS 11956&\cite{Kauffmann2015}&$169.2$&${\rm stellar\,abs.}$&$1.0$&$81.1$&$\geq176.6$&$(1)$\\
GASS 11845&\cite{Kauffmann2015}&$155.5$&${\rm stellar\,abs.}$&$1.1$&$95.7$&$167.9$&$(1)$\\
GASS 11824&\cite{Kauffmann2015}&$162.7$&${\rm H}\;\alpha$&$1.0$&$72.2$&$\geq196.7$&$-$\\
GASS 11808&\cite{Kauffmann2015}&$205.1$&${\rm stellar\,abs.}$&$1.2$&$122.3$&$\geq194.9$&$-$\\
GASS 11514&\cite{Kauffmann2015}&$183.3$&${\rm stellar\,abs.}$&$1.1$&$106.0$&$164.9$&$(1)$\\
GASS 11437&\cite{Kauffmann2015}&$113.1$&${\rm stellar\,abs.}$&$0.7$&$83.6$&$\geq183.0$&$(1)$\\
GASS 11386&\cite{Kauffmann2015}&$197.9$&${\rm stellar\,abs.}$&$1.2$&$96.3$&$148.1$&$-$\\
GASS 11349&\cite{Kauffmann2015}&$109.6$&${\rm stellar\,abs.}$&$0.7$&$98.5$&$\geq144.0$&$(1)$\\
GASS 11270&\cite{Kauffmann2015}&$169.2$&${\rm H}\;\alpha$&$1.0$&$50.6$&$54.5$&$-$\\
GASS 11223&\cite{Kauffmann2015}&$152.0$&${\rm stellar\,abs.}$&$1.1$&$117.5$&$\geq184.3$&$-$\\
GASS 11120&\cite{Kauffmann2015}&$116.1$&${\rm stellar\,abs.}$&$0.8$&$165.7$&$\geq197.8$&$(1)$\\
GASS 11087&\cite{Kauffmann2015}&$161.0$&${\rm stellar\,abs.}$&$1.0$&$116.1$&$191.8$&$(1)$\\
GASS 11019&\cite{Kauffmann2015}&$154.2$&${\rm stellar\,abs.}$&$0.9$&$87.6$&$\geq133.4$&$(1)$\\
GASS 10949&\cite{Kauffmann2015}&$112.2$&${\rm stellar\,abs.}$&$0.7$&$94.1$&$144.1$&$-$\\
GASS 10948&\cite{Kauffmann2015}&$110.5$&${\rm stellar\,abs.}$&$0.7$&$96.2$&$\geq142.5$&$(1)$\\
GASS 10943&\cite{Kauffmann2015}&$117.8$&${\rm stellar\,abs.}$&$0.7$&$105.7$&$152.3$&$(1)$\\
GASS 10942&\cite{Kauffmann2015}&$107.5$&${\rm stellar\,abs.}$&$0.6$&$42.0$&$\geq101.4$&$-$\\
GASS 10884&\cite{Kauffmann2015}&$110.1$&${\rm stellar\,abs.}$&$0.7$&$93.6$&$171.1$&$(1)$\\
GASS 10850&\cite{Kauffmann2015}&$152.0$&${\rm stellar\,abs.}$&$0.9$&$114.2$&$\geq204.7$&$(1)$\\
GASS 10841&\cite{Kauffmann2015}&$115.6$&${\rm stellar\,abs.}$&$0.7$&$165.9$&$166.0$&$-$\\
GASS 10831&\cite{Kauffmann2015}&$116.9$&${\rm H}\;\alpha$&$0.7$&$127.1$&$158.9$&$-$\\
GASS 10827&\cite{Kauffmann2015}&$128.9$&${\rm stellar\,abs.}$&$0.8$&$108.5$&$\geq166.2$&$-$\\
GASS 10813&\cite{Kauffmann2015}&$114.8$&${\rm stellar\,abs.}$&$0.7$&$117.6$&$149.5$&$-$\\
GASS 10447&\cite{Kauffmann2015}&$201.7$&${\rm stellar\,abs.}$&$1.2$&$57.1$&$\geq127.2$&$-$\\
GASS 10404&\cite{Kauffmann2015}&$154.6$&${\rm stellar\,abs.}$&$0.9$&$110.2$&$\geq166.0$&$-$\\
GASS 10358&\cite{Kauffmann2015}&$158.5$&${\rm stellar\,abs.}$&$1.0$&$52.2$&$\geq133.3$&$-$\\
GASS 10218&\cite{Kauffmann2015}&$198.7$&${\rm stellar\,abs.}$&$1.2$&$81.2$&$\geq117.5$&$(1)$\\
GASS 10019&\cite{Kauffmann2015}&$131.9$&${\rm stellar\,abs.}$&$0.8$&$117.2$&$\geq177.2$&$(1)$\\
WLM&\cite{Oh2015}&$1.0$&${\rm H}\,{\rm I}$&$<0.1$&$35.1$&$38.5$&$-$\\
UGC 8508&\cite{Oh2015}&$2.6$&${\rm H}\,{\rm I}$&$0.1$&$46.1$&$\geq46.1$&$-$\\
NGC 3738&\cite{Oh2015}&$4.9$&${\rm H}\,{\rm I}$&$0.1$&$125.6$&$\geq132.7$&$-$\\
NGC 2366&\cite{Oh2015}&$3.4$&${\rm H}\,{\rm I}$&$0.1$&$41.9$&$59.8$&$-$\\
NGC 1569&\cite{Oh2015}&$3.4$&${\rm H}\,{\rm I}$&$0.1$&$36.6$&$39.3$&$-$\\
LSB F564-V3&\cite{Oh2015}&$8.7$&${\rm H}\,{\rm I}$&$0.3$&$38.7$&$39.2$&$-$\\
IC 1613&\cite{Oh2015}&$0.7$&${\rm H}\,{\rm I}$&$<0.1$&$20.5$&$21.1$&$-$\\
IC 10&\cite{Oh2015}&$0.7$&${\rm H}\,{\rm I}$&$<0.1$&$-$&$\geq36.4$&$-$\\
Haro 36&\cite{Oh2015}&$9.3$&${\rm H}\,{\rm I}$&$0.3$&$37.6$&$\geq58.2$&$-$\\
Haro 29&\cite{Oh2015}&$5.9$&${\rm H}\,{\rm I}$&$0.2$&$34.4$&$43.5$&$-$\\
DDO 87&\cite{Oh2015}&$7.7$&${\rm H}\,{\rm I}$&$0.2$&$28.0$&$\geq56.6$&$-$\\
DDO 70&\cite{Oh2015}&$1.3$&${\rm H}\,{\rm I}$&$<0.1$&$43.9$&$\geq43.9$&$-$\\
DDO 53&\cite{Oh2015}&$3.6$&${\rm H}\,{\rm I}$&$0.1$&$29.2$&$\geq32.0$&$-$\\
DDO 52&\cite{Oh2015}&$10.3$&${\rm H}\,{\rm I}$&$0.3$&$42.6$&$\geq61.7$&$-$\\
DDO 50&\cite{Oh2015}&$3.4$&${\rm H}\,{\rm I}$&$0.1$&$31.2$&$38.8$&$-$\\
DDO 47&\cite{Oh2015}&$5.2$&${\rm H}\,{\rm I}$&$0.1$&$23.7$&$\geq64.7$&$-$\\
DDO 46&\cite{Oh2015}&$6.1$&${\rm H}\,{\rm I}$&$0.2$&$73.2$&$76.3$&$-$\\
DDO 43&\cite{Oh2015}&$7.8$&${\rm H}\,{\rm I}$&$0.2$&$31.5$&$\geq38.3$&$-$\\
DDO 216&\cite{Oh2015}&$1.1$&${\rm H}\,{\rm I}$&$<0.1$&$18.9$&$\geq18.9$&$-$\\
DDO 210&\cite{Oh2015}&$0.9$&${\rm H}\,{\rm I}$&$<0.1$&$-$&$\geq12.0$&$-$\\
DDO 168&\cite{Oh2015}&$4.3$&${\rm H}\,{\rm I}$&$0.1$&$57.5$&$61.9$&$-$\\
DDO 154&\cite{Oh2015}&$3.7$&${\rm H}\,{\rm I}$&$0.1$&$35.8$&$\geq51.1$&$-$\\
DDO 133&\cite{Oh2015}&$3.5$&${\rm H}\,{\rm I}$&$0.1$&$41.6$&$46.7$&$-$\\
DDO 126&\cite{Oh2015}&$4.9$&${\rm H}\,{\rm I}$&$0.1$&$30.7$&$38.7$&$-$\\
DDO 101&\cite{Oh2015}&$6.4$&${\rm H}\,{\rm I}$&$0.2$&$63.3$&$\geq64.9$&$-$\\
CVnIdwA&\cite{Oh2015}&$3.6$&${\rm H}\,{\rm I}$&$0.1$&$25.9$&$26.4$&$-$\\
UGC 2259&\cite{Adams2014}&$9.9$&${\rm H}\;\beta+{\rm O}\,{\rm III}$&$0.1$&$74.0$&$\geq93.4$&$-$\\
UGC 11707&\cite{Adams2014}&$15.0$&${\rm H}\;\beta+{\rm O}\,{\rm III}$&$0.2$&$51.6$&$\geq103.7$&$-$\\
NGC 959&\cite{Adams2014}&$9.9$&${\rm H}\;\beta+{\rm O}\,{\rm III}$&$0.1$&$78.6$&$\geq84.1$&$-$\\
NGC 5949&\cite{Adams2014}&$14.3$&${\rm H}\;\beta+{\rm O}\,{\rm III}$&$0.2$&$92.2$&$\geq111.2$&$-$\\
NGC 5204&\cite{Adams2014}&$3.2$&${\rm H}\;\beta+{\rm O}\,{\rm III}$&$<0.1$&$83.2$&$\geq89.4$&$-$\\
NGC 2976&\cite{Adams2014}&$3.6$&${\rm H}\;\beta+{\rm O}\,{\rm III}$&$0.1$&$74.8$&$\geq76.8$&$-$\\
NGC 2552&\cite{Adams2014}&$11.4$&${\rm H}\;\beta+{\rm O}\,{\rm III}$&$0.2$&$57.9$&$\geq96.1$&$-$\\
NGC 2366&\cite{Oh2011}&$3.4$&${\rm H}\,{\rm I}$&$0.2$&$43.2$&$\geq66.7$&$(2)$\\
M81 dw B&\cite{Oh2011}&$5.3$&${\rm H}\,{\rm I}$&$0.3$&$31.6$&$\geq39.5$&$-$\\
IC 2574&\cite{Oh2011}&$4.0$&${\rm H}\,{\rm I}$&$0.2$&$24.5$&$80.0$&$-$\\
Ho II&\cite{Oh2011}&$3.4$&${\rm H}\,{\rm I}$&$0.2$&$27.9$&$37.5$&$(2)$\\
Ho I&\cite{Oh2011}&$3.8$&${\rm H}\,{\rm I}$&$0.2$&$28.3$&$37.2$&$-$\\
DDO 53&\cite{Oh2011}&$3.6$&${\rm H}\,{\rm I}$&$0.2$&$32.4$&$32.5$&$(2)$\\
DDO 154&\cite{Oh2011}&$4.3$&${\rm H}\,{\rm I}$&$0.2$&$34.5$&$\geq50.0$&$(2)$\\
UGC 5750&\cite{KuziodeNaray2008}&$56.1$&${\rm H}\;\alpha$&$0.5$&$19.0$&$\geq73.4$&$-$\\
UGC 477&\cite{KuziodeNaray2008}&$35.5$&${\rm H}\;\alpha$&$0.3$&$41.3$&$\geq111.7$&$-$\\
UGC 4325&\cite{KuziodeNaray2008}&$10.1$&${\rm H}\;\alpha$&$0.1$&$82.4$&$\geq110.7$&$(2)$\\
UGC 191&\cite{KuziodeNaray2008}&$17.6$&${\rm H}\;\alpha$&$0.2$&$78.0$&$\geq97.2$&$-$\\
UGC 1551&\cite{KuziodeNaray2008}&$20.2$&${\rm H}\;\alpha$&$0.2$&$42.4$&$\geq82.5$&$-$\\
UGC 1281&\cite{KuziodeNaray2008}&$5.5$&${\rm H}\;\alpha$&$0.1$&$37.8$&$\geq45.8$&$-$\\
UGC 128&\cite{KuziodeNaray2008}&$60.0$&${\rm H}\;\alpha$&$0.6$&$46.1$&$144.9$&$-$\\
UGC 11820&\cite{KuziodeNaray2008}&$13.3$&${\rm H}\;\alpha$&$0.1$&$80.1$&$96.9$&$-$\\
NGC 959&\cite{KuziodeNaray2008}&$7.8$&${\rm H}\;\alpha$&$0.1$&$76.7$&$79.2$&$(2)$\\
NGC 7137&\cite{KuziodeNaray2008}&$22.5$&${\rm H}\;\alpha$&$0.2$&$53.4$&$\geq71.6$&$-$\\
NGC 4395&\cite{KuziodeNaray2008}&$3.5$&${\rm H}\;\alpha$&$<0.1$&$-$&$\geq32.7$&$-$\\
LSB F583-4&\cite{KuziodeNaray2008}&$49.0$&${\rm H}\;\alpha$&$0.5$&$42.1$&$\geq95.8$&$-$\\
LSB F583-1&\cite{KuziodeNaray2008}&$32.0$&${\rm H}\;\alpha$&$0.3$&$42.0$&$\geq72.4$&$-$\\
LSB F568-3&\cite{KuziodeNaray2008}&$77.0$&${\rm H}\;\alpha$&$0.7$&$41.4$&$\geq114.3$&$-$\\
LSB F563-V2&\cite{KuziodeNaray2008}&$61.0$&${\rm H}\;\alpha$&$0.6$&$52.9$&$\geq103.8$&$-$\\
LSB F563-1&\cite{KuziodeNaray2008}&$45.0$&${\rm H}\;\alpha$&$0.4$&$51.7$&$\geq146.4$&$-$\\
DDO 64&\cite{KuziodeNaray2008}&$6.1$&${\rm H}\;\alpha$&$0.1$&$59.8$&$\geq59.8$&$-$\\
NGC 925&\cite{deBlok2008}&$9.2$&${\rm H}\,{\rm I}$&$0.5$&$34.7$&$\geq119.9$&$-$\\
NGC 7793&\cite{deBlok2008}&$3.9$&${\rm H}\,{\rm I}$&$0.2$&$76.2$&$117.9$&$-$\\
NGC 7331&\cite{deBlok2008}&$14.7$&${\rm H}\,{\rm I}$&$0.8$&$253.2$&$268.1$&$-$\\
NGC 6946&\cite{deBlok2008}&$5.9$&${\rm H}\,{\rm I}$&$0.3$&$132.5$&$224.3$&$-$\\
NGC 5055&\cite{deBlok2008}&$10.1$&${\rm H}\,{\rm I}$&$0.6$&$185.3$&$211.6$&$-$\\
NGC 4826&\cite{deBlok2008}&$7.5$&${\rm H}\,{\rm I}$&$0.4$&$-166.7$&$180.2$&$(3)$\\
NGC 4736&\cite{deBlok2008}&$4.7$&${\rm H}\,{\rm I}$&$0.3$&$168.7$&$198.3$&$-$\\
NGC 3627&\cite{deBlok2008}&$9.3$&${\rm H}\,{\rm I}$&$0.5$&$178.0$&$207.1$&$-$\\
NGC 3621&\cite{deBlok2008}&$6.6$&${\rm H}\,{\rm I}$&$0.4$&$102.9$&$\geq159.2$&$-$\\
NGC 3521&\cite{deBlok2008}&$10.7$&${\rm H}\,{\rm I}$&$0.6$&$192.1$&$233.4$&$-$\\
NGC 3198&\cite{deBlok2008}&$13.8$&${\rm H}\,{\rm I}$&$0.8$&$76.7$&$158.7$&$-$\\
NGC 3031&\cite{deBlok2008}&$3.6$&${\rm H}\,{\rm I}$&$0.2$&$242.2$&$259.8$&$-$\\
NGC 2976&\cite{deBlok2008}&$3.6$&${\rm H}\,{\rm I}$&$0.2$&$75.0$&$\geq86.2$&$(2)$\\
NGC 2903&\cite{deBlok2008}&$8.9$&${\rm H}\,{\rm I}$&$0.5$&$120.1$&$215.5$&$-$\\
NGC 2841&\cite{deBlok2008}&$14.1$&${\rm H}\,{\rm I}$&$0.8$&$-$&$323.9$&$-$\\
NGC 2403&\cite{deBlok2008}&$3.2$&${\rm H}\,{\rm I}$&$0.2$&$97.4$&$143.9$&$-$\\
NGC 2366&\cite{deBlok2008}&$3.4$&${\rm H}\,{\rm I}$&$0.2$&$43.2$&$\geq66.7$&$(2)$\\
IC 2574&\cite{deBlok2008}&$4.0$&${\rm H}\,{\rm I}$&$0.2$&$20.6$&$\geq78.5$&$(2)$\\
DDO 154&\cite{deBlok2008}&$4.3$&${\rm H}\,{\rm I}$&$0.2$&$34.5$&$\geq50.0$&$(2)$\\
NGC 6822&\cite{deBlok2004}&$0.5$&${\rm H}\,{\rm I}$&$<0.1$&$41.8$&$\geq56.3$&$-$\\
UGC 8490&\cite{Swaters2003}&$20.5$&${\rm H}\,{\rm I}+{\rm H}\;\alpha$&$0.1$&$30.0$&$80.1$&$(2)$\\
UGC 731&\cite{Swaters2003}&$8.0$&${\rm H}\,{\rm I}+{\rm H}\;\alpha$&$<0.1$&$61.5$&$\geq74.0$&$(3)$\\
UGC 5721&\cite{Swaters2003}&$6.7$&${\rm H}\,{\rm I}+{\rm H}\;\alpha$&$<0.1$&$76.2$&$80.4$&$(2)$\\
UGC 4499&\cite{Swaters2003}&$13.0$&${\rm H}\,{\rm I}+{\rm H}\;\alpha$&$0.1$&$46.2$&$\geq74.2$&$(2)$\\
UGC 4325&\cite{Swaters2003}&$10.1$&${\rm H}\,{\rm I}+{\rm H}\;\alpha$&$0.1$&$77.0$&$104.6$&$(2)$\\
UGC 2259&\cite{Swaters2003}&$9.8$&${\rm H}\,{\rm I}+{\rm H}\;\alpha$&$0.1$&$78.8$&$93.7$&$(2)$\\
UGC 12732&\cite{Swaters2003}&$13.2$&${\rm H}\,{\rm I}+{\rm H}\;\alpha$&$0.1$&$45.8$&$\geq98.0$&$-$\\
UGC 11861&\cite{Swaters2003}&$25.1$&${\rm H}\,{\rm I}+{\rm H}\;\alpha$&$0.1$&$80.4$&$164.0$&$(2)$\\
UGC 11707&\cite{Swaters2003}&$15.9$&${\rm H}\,{\rm I}+{\rm H}\;\alpha$&$0.1$&$51.3$&$\geq99.9$&$(2)$\\
UGC 11557&\cite{Swaters2003}&$23.8$&${\rm H}\,{\rm I}+{\rm H}\;\alpha$&$0.1$&$35.1$&$\geq84.5$&$(2)$\\
LSB F574-1&\cite{Swaters2003}&$96.0$&${\rm H}\,{\rm I}+{\rm H}\;\alpha$&$0.5$&$57.9$&$\geq104.2$&$-$\\
LSB F568-V1&\cite{Swaters2003}&$80.0$&${\rm H}\,{\rm I}+{\rm H}\;\alpha$&$0.4$&$67.0$&$124.9$&$-$\\
LSB F568-3&\cite{Swaters2003}&$77.0$&${\rm H}\,{\rm I}+{\rm H}\;\alpha$&$0.4$&$32.5$&$111.2$&$(2)$\\
LSB F568-1&\cite{Swaters2003}&$85.0$&${\rm H}\,{\rm I}+{\rm H}\;\alpha$&$0.4$&$67.9$&$\geq130.7$&$-$\\
LSB F563-V2&\cite{Swaters2003}&$61.0$&${\rm H}\,{\rm I}+{\rm H}\;\alpha$&$0.3$&$87.7$&$\geq113.1$&$(2)$\\
UGC 711&\cite{deBlok2002}&$26.4$&${\rm H}\;\alpha$&$0.3$&$27.8$&$\geq91.6$&$-$\\
UGC 5750&\cite{deBlok2002}&$56.0$&${\rm H}\,{\rm I}+{\rm H}\;\alpha$&$0.5$&$20.0$&$\geq49.6$&$(2)$\\
UGC 5005&\cite{deBlok2002}&$52.0$&${\rm H}\,{\rm I}+{\rm H}\;\alpha$&$0.5$&$29.6$&$\geq100.0$&$-$\\
UGC 4325&\cite{deBlok2002}&$10.1$&${\rm H}\,{\rm I}+{\rm H}\;\alpha$&$0.1$&$73.8$&$\geq122.6$&$(2)$\\
UGC 4173&\cite{deBlok2002}&$16.8$&${\rm H}\,{\rm I}+{\rm H}\;\alpha$&$0.2$&$22.8$&$\geq57.0$&$-$\\
UGC 3371&\cite{deBlok2002}&$12.8$&${\rm H}\,{\rm I}+{\rm H}\;\alpha$&$0.1$&$33.9$&$\geq85.7$&$-$\\
UGC 3137&\cite{deBlok2002}&$18.4$&${\rm H}\,{\rm I}+{\rm H}\;\alpha$&$0.2$&$45.6$&$106.9$&$-$\\
UGC 1281&\cite{deBlok2002}&$5.5$&${\rm H}\,{\rm I}+{\rm H}\;\alpha$&$0.1$&$37.7$&$\geq56.9$&$(2)$\\
UGC 1230&\cite{deBlok2002}&$51.0$&${\rm H}\,{\rm I}+{\rm H}\;\alpha$&$0.5$&$50.5$&$112.7$&$-$\\
UGC 10310&\cite{deBlok2002}&$15.6$&${\rm H}\,{\rm I}+{\rm H}\;\alpha$&$0.1$&$38.9$&$\geq75.0$&$(2)$\\
NGC 5023&\cite{deBlok2002}&$4.8$&${\rm H}\,{\rm I}+{\rm H}\;\alpha$&$<0.1$&$64.9$&$\geq84.4$&$-$\\
NGC 4455&\cite{deBlok2002}&$6.8$&${\rm H}\,{\rm I}+{\rm H}\;\alpha$&$0.1$&$44.9$&$\geq64.4$&$-$\\
NGC 4395&\cite{deBlok2002}&$3.5$&${\rm H}\,{\rm I}+{\rm H}\;\alpha$&$<0.1$&$57.0$&$\geq84.2$&$(2)$\\
NGC 3274&\cite{deBlok2002}&$6.7$&${\rm H}\,{\rm I}+{\rm H}\;\alpha$&$0.1$&$82.5$&$82.6$&$(2)$\\
NGC 2366&\cite{deBlok2002}&$3.4$&${\rm H}\,{\rm I}+{\rm H}\;\alpha$&$<0.1$&$54.0$&$55.5$&$(2)$\\
NGC 1560&\cite{deBlok2002}&$3.0$&${\rm H}\,{\rm I}+{\rm H}\;\alpha$&$<0.1$&$42.6$&$\geq77.5$&$-$\\
NGC 100&\cite{deBlok2002}&$11.2$&${\rm H}\;\alpha$&$0.1$&$60.0$&$\geq91.2$&$-$\\
LSB F563-1&\cite{deBlok2002}&$45.0$&${\rm H}\,{\rm I}+{\rm H}\;\alpha$&$0.4$&$57.1$&$\geq114.1$&$(2)$\\
IC 2233&\cite{deBlok2002}&$10.5$&${\rm H}\,{\rm I}+{\rm H}\;\alpha$&$0.1$&$40.8$&$\geq92.8$&$-$\\
DDO 64&\cite{deBlok2002}&$6.1$&${\rm H}\,{\rm I}+{\rm H}\;\alpha$&$0.1$&$46.3$&$\geq46.9$&$(2)$\\
DDO 52&\cite{deBlok2002}&$5.3$&${\rm H}\,{\rm I}+{\rm H}\;\alpha$&$0.1$&$43.1$&$\geq50.0$&$(2)$\\
DDO 47&\cite{deBlok2002}&$4.0$&${\rm H}\,{\rm I}+{\rm H}\;\alpha$&$<0.1$&$44.0$&$\geq67.0$&$(2)$\\
DDO 189&\cite{deBlok2002}&$12.6$&${\rm H}\,{\rm I}+{\rm H}\;\alpha$&$0.1$&$47.6$&$\geq65.7$&$-$\\
DDO 185&\cite{deBlok2002}&$5.1$&${\rm H}\,{\rm I}+{\rm H}\;\alpha$&$<0.1$&$43.3$&$\geq49.6$&$-$\\
UGC 6614&\cite{deBlok2001}&$85.0$&${\rm H}\;\alpha$&$0.6$&$120.2$&$\geq205.2$&$-$\\
UGC 5750&\cite{deBlok2001}&$56.0$&${\rm H}\;\alpha$&$0.4$&$23.2$&$\geq78.9$&$(2)$\\
UGC 4115&\cite{deBlok2001}&$3.2$&${\rm H}\,{\rm I}+{\rm H}\;\alpha$&$<0.1$&$-$&$\geq39.8$&$-$\\
UGC 11819&\cite{deBlok2001}&$60.0$&${\rm H}\,{\rm I}+{\rm H}\;\alpha$&$0.4$&$73.0$&$\geq154.7$&$-$\\
UGC 11748&\cite{deBlok2001}&$73.0$&${\rm H}\,{\rm I}+{\rm H}\;\alpha$&$0.5$&$198.9$&$250.0$&$-$\\
UGC 11648&\cite{deBlok2001}&$48.0$&${\rm H}\,{\rm I}+{\rm H}\;\alpha$&$0.3$&$74.3$&$\geq144.6$&$-$\\
UGC 11616&\cite{deBlok2001}&$73.0$&${\rm H}\,{\rm I}+{\rm H}\;\alpha$&$0.5$&$87.8$&$\geq142.8$&$-$\\
UGC 11583&\cite{deBlok2001}&$5.0$&${\rm H}\,{\rm I}+{\rm H}\;\alpha$&$<0.1$&$35.6$&$\geq35.6$&$-$\\
UGC 11557&\cite{deBlok2001}&$22.0$&${\rm H}\,{\rm I}+{\rm H}\;\alpha$&$0.2$&$34.5$&$\geq80.4$&$(2)$\\
UGC 11454&\cite{deBlok2001}&$91.0$&${\rm H}\,{\rm I}+{\rm H}\;\alpha$&$0.7$&$85.8$&$\geq152.2$&$-$\\
LSB F730-V1&\cite{deBlok2001}&$144.0$&${\rm H}\;\alpha$&$1.0$&$87.3$&$\geq145.3$&$-$\\
LSB F583-4&\cite{deBlok2001}&$49.0$&${\rm H}\,{\rm I}+{\rm H}\;\alpha$&$0.4$&$44.5$&$\geq69.9$&$(2)$\\
LSB F583-1&\cite{deBlok2001}&$32.0$&${\rm H}\,{\rm I}+{\rm H}\;\alpha$&$0.2$&$40.7$&$\geq86.9$&$(2)$\\
LSB F579-V1&\cite{deBlok2001}&$85.0$&${\rm H}\,{\rm I}+{\rm H}\;\alpha$&$0.6$&$93.5$&$\geq114.4$&$-$\\
LSB F571-8&\cite{deBlok2001}&$48.0$&${\rm H}\,{\rm I}+{\rm H}\;\alpha$&$0.3$&$68.1$&$\geq143.9$&$-$\\
LSB F568-3&\cite{deBlok2001}&$77.0$&${\rm H}\,{\rm I}+{\rm H}\;\alpha$&$0.6$&$41.0$&$\geq101.1$&$(2)$\\
LSB F563-1&\cite{deBlok2001}&$23.6$&${\rm H}\,{\rm I}+{\rm H}\;\alpha$&$0.2$&$84.0$&$112.4$&$(2)$\\
ESO 4880490&\cite{deBlok2001}&$22.0$&${\rm H}\;\alpha$&$0.2$&$62.3$&$\geq97.1$&$-$\\
ESO 4250180&\cite{deBlok2001}&$86.0$&${\rm H}\;\alpha$&$0.6$&$-$&$\geq144.5$&$-$\\
ESO 3050090&\cite{deBlok2001}&$11.0$&${\rm H}\;\alpha$&$0.1$&$34.6$&$\geq54.6$&$-$\\
ESO 3020120&\cite{deBlok2001}&$69.0$&${\rm H}\;\alpha$&$0.5$&$48.3$&$\geq86.3$&$-$\\
ESO 2060140&\cite{deBlok2001}&$60.0$&${\rm H}\;\alpha$&$0.4$&$83.6$&$\geq118.0$&$-$\\
ESO 1870510&\cite{deBlok2001}&$18.0$&${\rm H}\;\alpha$&$0.1$&$36.3$&$\geq39.9$&$-$\\
ESO 1200211&\cite{deBlok2001}&$15.0$&${\rm H}\;\alpha$&$0.1$&$21.8$&$\geq25.4$&$-$\\
ESO 0840411&\cite{deBlok2001}&$80.0$&${\rm H}\;\alpha$&$0.6$&$19.7$&$\geq61.3$&$-$\\
ESO 0140040&\cite{deBlok2001}&$212.0$&${\rm H}\;\alpha$&$1.5$&$120.3$&$\geq272.7$&$-$\\

\end{longtable}

\label{lastpage}

\end{document}